\documentstyle[aps,pre,epsfig,titlepage]{revtex}

\def\sequenceone{{\vec a}}
\def\sequencetwo{{\vec b}}
\def\letterone{a}
\def\lettertwo{b}
\def\score{h}
\def\localscore{S}
\def\smatrix{s}
\def\maxscore{\Sigma}
\def\islandscore{\sigma}
\def\timevar{t}
\def\width{W}
\def\binsize{{n_{\mathrm max}}}
\def\disorder{\eta}
\def\occupation{n}
\def\hopnumber{j}
\def\totalhops{J}
\def\genfunc{Q}

\def\eigenval{\rho}
\def\gfgeneral{Z}

\draft

\title{Asymmetric exclusion process and extremal statistics of random
sequences}
\author{R.~Bundschuh}
\address{
Department of Physics, University of California at San Diego,
La Jolla, CA  92093-0319, U.S.A.}
\date{24. November 1999}

\begin{document}

\maketitle

\thispagestyle{empty}

\begin{abstract}
An exact mapping is established between sequence alignment, one of the
most commonly used tools of computational biology, and the asymmetric
exclusion process, one of the few exactly solvable models of
nonequilibrium physics.  The statistical significance of sequence
alignments is characterized through studying the total hopping current
of the discrete time and space version of the asymmetric exclusion
process.

\end{abstract}

\section{Introduction}

Sequence alignment is one of the most commonly used computational
tools of molecular biology. Its applications range from the
identification of the function of newly sequenced genes to the
construction of phylogenic trees~\cite{wate94,dool96}. Beyond its
practical importance, it is one of the simplest model systems for
pattern matching.  In computational biology, sequences are routinely
compared via a transfer matrix algorithm to find the ``optimal''
alignment. Recently, it has been noted that this transfer matrix
algorithm is the same as the one used to calculate the partition
function or optimal energy of a directed polymer in a random
medium~\cite{hwa96}. This problem is known to belong to the
universality class of surface growth as described by the
Kardar-Parisi-Zhang (KPZ) equation~\cite{kard86}. From the assignment
of sequence alignment to the KPZ universality class various {\em
scaling laws} characterizing sequence alignment have been deduced.
They have been used in order to answer questions of practical
importance to sequence alignment, e.g., the optimal choice of
alignment parameters~\cite{dras98a,hwa98,dras98b}.  But there are also
{\em non universal} features which are of great importance for
practical applications. They cannot be extracted from the knowledge of
the universality class alone, but have to be evaluated by a
microscopic study taking into account all the details of the given
sequence alignment algorithm. In this paper, we will perform such a
study for a certain choice of parameters for which sequence alignment
maps {\em exactly} onto the asymmetric exclusion
process~\cite{krug91a,derr98a}, which is the best studied
nonequilibrium system of the KPZ universality class, equivalent also
to the six vertex model~\cite{kand90,gwa92}.

We will apply this mapping to address the central question in the
biological application of sequence alignment, namely the assessment of
alignment significance: The problem is that an ``optimal'' alignment,
i.e., the best possible alignment of two given sequences according to
some scoring function, does not necessarily reflect sequence homology.
A sequence alignment algorithm will produce an ``optimal'' alignment
for any pair of sequences, including randomly chosen ones. The
important question is whether the alignment produced reflects an
underlying similarity of the two sequences compared.  A common way to
address this question is to evaluate the probability of getting a
certain alignment score by chance. This requires the knowledge of the
distribution of alignment scores for random sequences. This
distribution turns out to obey a universal (Gumbel) form with two
non-universal parameters.  In this paper, we will derive the Gumbel
distribution and characterize some of its properties by relating them
to the corresponding asymmetric exclusion process.  In particular, we
show how the tail of this distribution can be obtained from the
generating function for the total number of hopped particles.  The
latter is also the generating function of the average surface height
in the equivalent surface growth formulation of the asymmetric
exclusion process. This important quantity has been calculated for the
case of continuous time and continuous space using the replica
trick~\cite{kard85} a long time ago. More recently, it has been
obtained for the case of continuous time and discrete space in the
scaling regime~\cite{derr98b}. Here, we will calculate this quantity
in discrete time and discrete space as necessary for the mapping to
sequence alignment, in the asymptotic large size limit which is {\em
beyond} the scaling limit.  Our calculation does not make use of the
replica trick and leads to a very simple closed form expression. It
explicitly contains the anomalous $t^{1/3}$ scaling of the surface
height fluctuations of KPZ surface growth in one dimension. We use
this generating function to give an explicit expression for the
significance of sequence alignments.

The paper is organized as follows: First, we will give a
self-contained introduction to sequence alignment in
Sec.~\ref{sec_seqalign}. This familiarizes the reader with the
sequence alignment algorithm and gives us a chance to develop the
notations to be used later.  In Sec.~\ref{sec_localtoglobal}, we will
reduce the problem of assessing the statistical significance of the
widely used {\em local} alignment to a quantity defined in terms of
the simpler {\em global} alignment. Readers more interested in the
properties of the discrete asymmetric exclusion process can skip these
two sections and go directly to Sec.~\ref{sec_maptoasep}, which
describes the simplest version of the global alignment problem. Here,
the exact mapping to the asymmetric exclusion process in discrete time
and space with sublattice-parallel updating is described.
Sec.~\ref{sec_evalgenfunc} is devoted to the calculation of the
generating function of interest for the asymmetric exclusion process.
In Sec.~\ref{sec_implications}, we discuss the result obtained, apply
it to the assessment of alignment significance, and verify the
analytical predictions numerically. In Sec.~\ref{sec_generalize}, we
consider more general scoring systems and map them onto generalized
asymmetric exclusion processes.  The final section gives a short
summary of the paper and points towards several future directions. A
number of technical details are given in the appendices.

\section{Review of Sequence Alignment}\label{sec_seqalign}

\subsection{Gapless Alignment}

Sequence alignment algorithms come in different levels of
sophistication.  The simplest alignment algorithm is {\em gapless}
alignment. It is not only extremely fast but also very well understood
theoretically. Thus, it has been very widely used, e.g., in its
implementation of the program BLAST~\cite{alts90}.

Gapless alignment looks for similarities between two sequences
$\sequenceone = \{\letterone_1 \letterone_2 \ldots \letterone_M\}$,
and $\sequencetwo=\{\lettertwo_1 \lettertwo_2 \ldots \lettertwo_{N}\}$
of length $M$ and $N\sim M$ respectively. The letters $\letterone_i$
and $\lettertwo_j$ are taken from an alphabet of size $c$. This may be
the four letter alphabet $\{A,C,G,T\}$ of DNA sequences or the twenty
letter alphabet of protein sequences with the letters distributed
according to the natural frequencies of the twenty amino acids. A
local gapless alignment ${\cal A}$ of these two sequences consists of
a substring $\letterone_{i-\ell+1}\dots \letterone_{i-1}\letterone_i$
of length $\ell$ of sequence $\sequenceone$ and a substring
$\lettertwo_{j-\ell+1}\dots \lettertwo_{j-1}\lettertwo_j$ of sequence
$\sequencetwo$ of the same length. Each such alignment is assigned a
score
\begin{equation}
\localscore[{\cal A}]=\localscore(i,j,\ell)=\sum_{k=0}^{\ell-1}
\smatrix_{\letterone_{i-k},
\lettertwo_{j-k}},
\end{equation}
where $\smatrix_{\letterone,\lettertwo}$ is some given ``scoring
matrix'' measuring the mutual degree of similarity of the different
letters of the alphabet.  A simple example of such a scoring matrix is
the match--mismatch matrix
\begin{equation}\label{eq_idmatrix}
\smatrix_{\letterone,\lettertwo}=
\left\{\begin{array}{ll}1&\letterone=\lettertwo\\
-\mu&\letterone\not=\lettertwo\end{array}\right.
\end{equation}
which is used for DNA sequence comparisons~\cite{need70}. For protein
sequences, the more complicated $20\times20$ PAM~\cite{dayh78} or
BLOSUM matrices~\cite{heni92} are used to account for the variable
degrees of similarity (e.g., hydrophobicity, size) among the $20$
amino acids. The computational task is to find the $i$, $j$, and
$\ell$ which give the {\em highest} total score
\begin{equation}\label{eq_maxscore}
\maxscore\equiv\max_{\cal A} \localscore[{\cal A}]
\end{equation}
for a given scoring matrix $\smatrix_{\letterone,\lettertwo}$.

The optimization task called for in gapless alignment can be easily
accomplished by introducing an auxiliary quantity,
$\localscore_{i,j}$, which is the optimal score of the above
consecutive subsequences ending at $(i,j)$ (optimized over $\ell$.)
It can be conveniently calculated in $O(N^2)$ instead of the expected
$O(N^3)$ steps using the transfer matrix algorithm
\begin{equation}\label{eq_gaplessevolv0}
\localscore_{i,j}=
\max\{\localscore_{i-1,j-1}+\smatrix_{\letterone_i,\lettertwo_j},0\},
\end{equation}
with the initial condition $\localscore_{0,k}=0=\localscore_{k,0}$.
This recursion equation reflects that for a given $(i,j)$ the optimal
$\ell$ is either zero or larger than zero. If the optimal $\ell$ is
zero the corresponding score is zero as well. If the optimal $\ell$ is
at least one, the pair $(\letterone_i,\lettertwo_j)$ certainly belongs
to the optimal alignment together with whatever has been chosen to be
optimal up to the point $(i-1,j-1)$. Eq.~(\ref{eq_gaplessevolv0}) is
basically a random walk with increments
$\smatrix_{\letterone,\lettertwo}$ which is cut off if it falls below
zero. The global optimal score is obtained as
\begin{equation}\label{eq_Sigma}
\maxscore = \max_{1\le i \le M, 1 \le j \le N} \localscore_{i,j}.
\end{equation}

In order to characterize the statistical significance of the
alignment, it is necessary to know the distribution of $\maxscore$ for
gapless alignments of two {\em random} sequences, whose elements
$\letterone_k$'s are generated independently from the same frequencies
$p_\letterone$ as the query sequences, and scored with the same matrix
$\smatrix_{\letterone,\lettertwo}$. This distribution of $\maxscore$
has been worked out rigorously~\cite{karl92,karl93}. For suitable
scoring parameters, it is a Gumbel or extreme value distribution given
by
\begin{equation}\label{eq_gumbeldist}
\Pr\{\maxscore<\localscore\}=\exp(-\kappa e^{-\lambda\localscore}).
\end{equation}
This distribution is characterized by the two parameters $\lambda$ and
$\kappa$ with $\lambda$ giving the tail of the distribution and
$\lambda^{-1}\log\kappa$ describing the mode. For gapless alignment,
these non universal parameters can be explicitly
calculated~\cite{karl92,karl93} from the scoring matrix
$\smatrix_{\letterone,\lettertwo}$ and the letter frequencies
$p_\letterone$. For example, $\lambda$ is the unique positive solution
of the equation
\begin{equation}\label{eq_lambdacond}
\langle \exp(\lambda \smatrix)\rangle\equiv
\sum_{\letterone,\lettertwo}p_\letterone p_\lettertwo
\exp(\lambda \smatrix_{\letterone,\lettertwo})=1.
\end{equation}
The other parameter $\kappa$ is given by $\kappa=KMN$, where $K$ is a
more complicated function of the scoring matrix and the letter
frequencies. Instead of reviewing the full derivation of the
distribution Eq.~(\ref{eq_gumbeldist}) and its parameters, we will
below give some heuristic arguments which yield the known
result. These can later be generalized to the more relevant case of
alignment with gaps.

For random sequences, one can take $j=i$ in (\ref{eq_gaplessevolv0})
without loss of generality. Eq.~(\ref{eq_gaplessevolv0}) then becomes
a discrete Langevin equation, with
\begin{equation}
\localscore_{i,i} \equiv\localscore(i)=\max\{\localscore(i-1)+\smatrix(i),0\},
\label{eq_gaplessevolv}
\end{equation}
where the ``noise'' $\smatrix(i)\equiv
\smatrix_{\letterone,\lettertwo}$ is uncorrelated and given by the
distribution
\begin{equation}\label{eq_scoredist}
\Pr\{\smatrix_i>\smatrix\}=
\sum_{\{\letterone,\lettertwo|\smatrix_{\letterone,\lettertwo}>\smatrix\}}
p_\letterone p_\lettertwo.
\end{equation}

The dynamics of the evolution equation (\ref{eq_gaplessevolv}) can be
in two distinct phases. The quantity which distinguishes these two
phases is the expected local similarity score
\begin{equation}\label{eq_averscoreneg}
\langle \smatrix\rangle\equiv\sum_{\letterone,\lettertwo}
p_\letterone p_\lettertwo \smatrix_{\letterone,\lettertwo}.
\end{equation}
If it is positive, the score $\localscore(i)$ will increase on
average. After a while, it becomes positive enough that the maximum in
Eq.~(\ref{eq_gaplessevolv}) will never be given by the zero
option. This option could thus be omitted which corresponds to {\em
global} gapless alignment. The dynamics is then a random walk
$\localscore(i)=\localscore(i-1)+\smatrix(i)$ with an average upward
drift $\langle \smatrix\rangle$.  The maximal score will be close to
the end of the sequences and will be given by $\maxscore\approx
N\cdot\langle \smatrix\rangle$. Since it is linear in the length of
the sequences, this is called the {\em linear phase} of local
alignment. It is obviously not suited to identify matches of {\em
subsequences}, and the distribution of the maximal score $\maxscore$
is not an extreme value distribution. (It is just a sum of many
independent local scores $\smatrix(i)$ and therefore obeys a Gaussian
distribution according to the central limit theorem.)

\begin{figure}[ht]
\begin{center}
\epsfig{figure=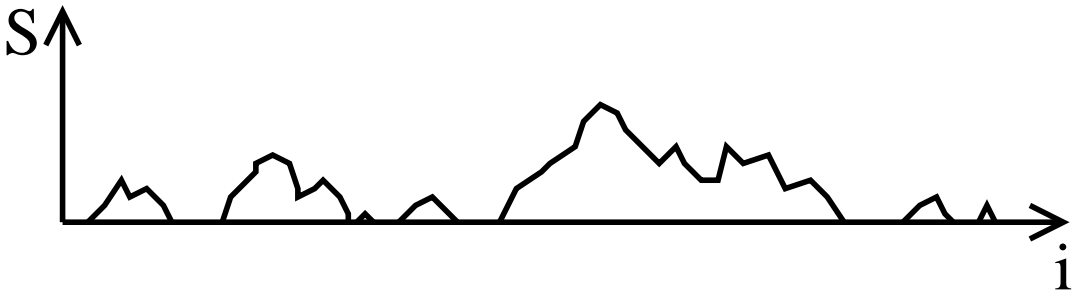,width=0.8\columnwidth}
\caption{Sketch of the total score as a function of sequence
position in gapless local alignment.}
\label{fig_gaplessscore}
\end{center}
\end{figure}

The situation is dramatically different if $\langle\smatrix\rangle$ is
negative.  In this case the dynamics is qualitatively as follows: The
score $\localscore(i)$ starts at zero. If the next local score
$\smatrix(i+1)$ is negative --- which is the more typical case in this
regime --- then $\localscore$ remains zero. But if the next local
score is positive, then $\localscore$ will increase by that
amount. Once it is positive, $\localscore(i)$ performs a random walk
with independent increments $\smatrix(i)$.  Since
$\langle\smatrix\rangle$ is negative, there is a {\em negative
drift\/} which forces $\localscore(i)$ to eventually return to
zero. After it is reset to zero, the whole process starts over again.
The qualitative ``temporal'' behavior of the score $\localscore(i)$ is
depicted in Fig.~\ref{fig_gaplessscore}.

{From} the figure, it is clear that the score landscape can be divided
into a series of ``{\em islands\/}'' of positive scores, separated by
``oceans'' where $\localscore=0$.  Each such island originates from a
single jump out of the zero-score state and terminates when the
zero-score state is reached again.  Since each of these islands
depends on a different subset of independent random numbers
$\smatrix(i)$, the islands are {\em statistically independent} of each
other.  If we let the maximal score of the $k^{\rm th}$ island be
$\islandscore_k$, then these $\islandscore_k$ are independent random
variables.  Calculating the probability for the maximum score
$\islandscore_k$ of an island of length $L$ in a saddle point
approximation and optimizing over the length $L$ of the islands, we
asymptotically obtain a Poisson distribution
\begin{equation}\label{eq_pislandpeak}
\Pr(\islandscore_k>\islandscore)\approx C_*e^{-\lambda\islandscore}
\end{equation}
for the maximal island scores $\islandscore_k$ (see
App.~\ref{app_gllambda}.)  The parameter $\lambda$ which gives the
typical scale of the maximal island score is given by the
drift-diffusion balance of the underlying Brownian process. If the
local scores $\smatrix(i)$ were Gaussian variables with average $v<0$
and variance $D$, this drift-diffusion balance would yield
\begin{equation}\label{eq_diffusivelambda}
\lambda=2\frac{|v|}{D}.
\end{equation}
For an arbitrary discrete or continuous distribution of the local
scores $\smatrix(i)$, it turns out to be given by the more general
condition~(\ref{eq_lambdacond}), which reduces to
Eq.~(\ref{eq_diffusivelambda}) in the limit
$\langle\smatrix\rangle\to0^-$ where the central limit theorem takes
hold.

Since the global optimal score $\maxscore$ can be expressed by the maximal
island scores as
\begin{equation}\label{eq_sigmamax}
\maxscore = \max_k \{\islandscore_k\},
\end{equation}
the distribution of $\maxscore$ can be calculated from the
distribution of the $\islandscore_k$. The connection is covered by the
theory of extremal statistics as developed by
Gumbel~\cite{gumb58,gala78}. In the case of a large number $K_*\sim N$
of independent island peak scores each of which asymptotically obeys
the Poisson distribution Eq.~(\ref{eq_pislandpeak}), the connection is
especially simple and we get
\begin{eqnarray}\label{eq_derivegumbel}
\Pr\{\maxscore<\localscore\}&=&
\Pr\{\max\{\islandscore_1,\ldots,\islandscore_{K_*}\}<\localscore\}
=\Pr\{\islandscore_1<\localscore\}^{K_*}\\\nonumber
&=&(1-C_*e^{-\lambda\localscore})^{K_*}
\approx[\exp(-C_*e^{-\lambda\localscore})]^{K_*}
=\exp(-\kappa e^{-\lambda\localscore})
\end{eqnarray}
with $\kappa\equiv C_*K_*$, i.e., the parameter $\lambda$ of the
island peak score distribution Eq.~(\ref{eq_pislandpeak}) is the same
as the parameter $\lambda$ in the Gumbel distribution
Eq.~(\ref{eq_gumbeldist}) of the maximal alignment scores.

\subsection{Alignment with Gaps}

In order to detect weak similarities between sequences separated by a
large evolutionary distance, ``gaps'' have to be allowed within an
alignment to compensate for insertions or deletions occurred during
the course of evolution~\cite{pear91}.  Here, we will specifically
consider Smith-Waterman local alignment~\cite{smit81}.  In this case,
a possible alignment ${\cal A}$ still consists of two substrings of
the two original sequences $\sequenceone$ and $\sequencetwo$.  But
now, these subsequences may have different lengths, since gaps may be
inserted in the alignment. For example the two subsequences {\tt
GATGC} and {\tt GCTC} may be aligned as {\tt GATGC} and {\tt GCT-C}
using one gap. Each such alignment ${\cal A}$ is assigned a score
according to
\begin{equation}
\localscore[{\cal A}]=
\sum_{(\letterone,\lettertwo)\in{\cal A}}
\smatrix_{\letterone,\lettertwo}-\delta N_{\mathrm g}
\end{equation}
where the sum is taken over all pairs of aligned letters, $N_{\mathrm
g}$ is the total number of gaps in the alignment, and $\delta$ is an
additional scoring parameter, the ``gap cost''. In practice more
complicated gap scores may be used, but we will concentrate on this
version.

\begin{figure}[htbp]
%\narrowtext
\begin{center}
\epsfig{figure=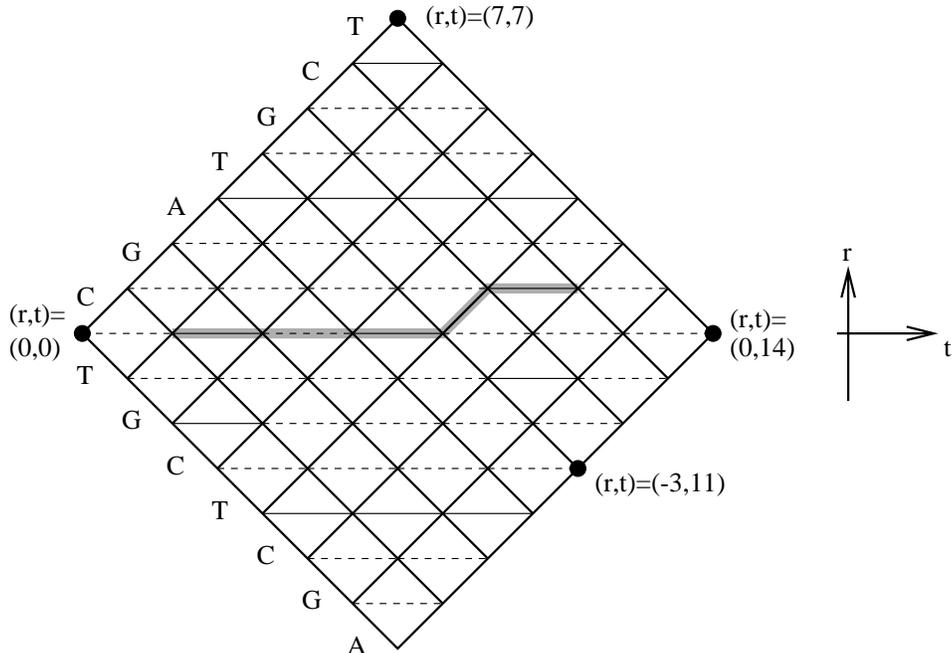,width=0.7\columnwidth}
%FIGSPACE:60mm
\caption{Local alignment of two sequences \protect$CGATGCT$ and
\protect$TGCTCGA$ represented as a directed path on the alignment
lattice: the diagonal bonds correspond to gaps in the alignment. The
horizontal bonds represent aligned pairs. Alignments of identical
letters (matches) are shown as solid lines; alignments of different
letters (mismatches) are shown dashed. The highlighted alignment path
\protect$r(\timevar)$ corresponds to one possible alignment of two
subsequences, \protect$GATGC$ to \protect$GCT\!-\!C$. This path
contains one gap.  It is also shown how the coordinates $r$ and $\timevar$
are used to identify the nodes of the lattice.}
\label{fig_alpath}
\end{center}
\end{figure}

The task of local alignment is again to find the alignment ${\cal A}$
with the highest score as in Eq.~(\ref{eq_maxscore}), in this enlarged
class of possible alignments. This can be very efficiently done by a
transfer matrix method which becomes obvious in the alignment path
representation~\cite{need70}. In this representation, the two sequences
to be compared are written on the edges of a square lattice as the one
shown in Fig.~\ref{fig_alpath} where we chose for simplicity
$N=M$. Each directed path on this lattice represents one
possible alignment. The score of this alignment is the sum over the
local scores of the traversed bonds. Diagonal bonds correspond
to gaps and carry the score $-\delta$. Horizontal bonds are assigned
the similarity scores
\begin{equation}
\smatrix(r,\timevar)\equiv \smatrix_{\letterone_i,\lettertwo_j}
\end{equation}
where $\letterone_i$ and $\lettertwo_j$ are the letters of the two
sequences belonging to the position $(r,\timevar)=(i-j,i+j-1)$ as
shown in Fig.~\ref{fig_alpath}.

If we were interested in finding the highest scoring {\em global}
alignment of the two sequences $\sequenceone$ and $\sequencetwo$, this
corresponds to finding the best scoring path connecting the beginning
$(0,0)$ with the end $(0,2N)$ of the lattice. To find this path
effectively, we define the auxiliary quantity $\score(r,\timevar)$ to be
the score of the best path ending in the lattice point
$(r,\timevar)$. This quantity can be calculated by the
Needleman-Wunsch transfer matrix algorithm~\cite{need70}
\begin{equation}\label{eq_globrecurs}
\score(r,\timevar+1)=
\max\{\score(r,\timevar-1)+\smatrix(r,\timevar),\
\score(r+1,\timevar)-\delta,
\score(r-1,\timevar)-\delta\}.
\end{equation}
This is easily recognized~\cite{hwa96} as the algorithm used to
calculate the zero temperature configuration and energy of a directed
polymer in a random potential given by the local scores
$\smatrix(r,\timevar)$. The scores $\score(r,\timevar)$ represent the
(negative) energy of the optimally chosen polymer configuration ending
in the point $(r,\timevar)$. Alternatively, the $\score(r,\timevar)$
can also be interpreted as the spatial height profile of a growing
surface through the well known relation between the directed polymer
and the KPZ equation.

\begin{figure}[ht]
\begin{center}
\epsfig{figure=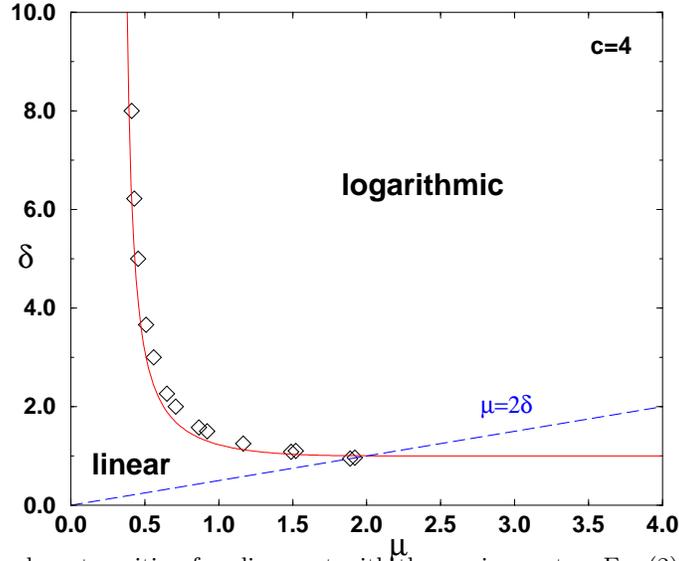,angle=270,width=0.5\columnwidth}
\caption{Loci of the log-linear phase transition for alignment with
the scoring system Eq.~(\protect\ref{eq_idmatrix}) for an alphabet of
\protect$c=4$ letters in terms of the mismatch cost \protect$\mu$ and
the gap cost \protect$\delta$. Useful alignments can only be
obtained in the logarithmic phase above the phase transition line. The
diamonds are numerically estimated points on the phase transition
line; the solid line is the approximate locus calculated
in~\protect\cite{bund99}. Below the dashed line the alignments do not
depend on the mismatch cost \protect$\mu$ any more and the phase
transition line is known to be strictly horizontal.}
\label{fig_ptline}
\end{center}
\end{figure}

If we are interested in {\em local} alignments, we can use the same
trick as in the gapless case~(\ref{eq_gaplessevolv0}). Cutting off
unfavorable scores by adding the choice of zero in the maximum of
Eq.~(\ref{eq_globrecurs}) leads to the Smith-Waterman
algorithm~\cite{smit81}
\begin{equation}\label{eq_swrecursion}
\localscore(r,\timevar+1)=\max\left\{\begin{array}{l}
\localscore(r,\timevar-1)+\smatrix(r,\timevar)\\
\localscore(r+1,\timevar)-\delta\\\localscore(r-1,\timevar)-\delta\\0
\end{array}\right\}.
\end{equation}
The score of the best local alignment is then given by
\begin{equation}
\maxscore=\max_{r,\timevar}\localscore(r,\timevar).
\end{equation}
In the presence of gaps, we can still distinguish a linear and a
logarithmic phase. If the global alignment score tends to grow, the
zero option of the local alignment algorithm does not play any role.
We effectively revert to global alignment and get a maximum score
which is linear in the length of the sequences. Contrary to gapless
alignment, it is not enough to have a negative expectation value of
the local scores $\langle\smatrix\rangle$ in order to prevent
this. This is due to the fact that the alignment algorithm uses gaps
to connect random stretches of good matches to optimize the score. The
average score grows by a gap dependent amount
$u(\{\smatrix_{\letterone,\lettertwo}\},\delta)$ faster compared to
the expectation value $\langle\smatrix\rangle$.  The log-linear
transition occurs now at
$u(\{\smatrix_{\letterone,\lettertwo}\},\delta_{\mathrm
c})+\langle\smatrix\rangle=0$. For the simple scoring system
Eq.~(\ref{eq_idmatrix}) this corresponds to a line $\delta_{\mathrm
c}(\mu)$ in the two dimensional space of the parameters $\mu$ and
$\delta$ shown in Fig.~\ref{fig_ptline}. Even for this simple scoring
system, the loci of the phase transition are only known
approximately~\cite{bund99}; for more complicated scoring systems,
only numerical results are available.

If the parameters are chosen such that $u+\langle\smatrix\rangle<0$,
i.e., such that the expected global alignment score drifts downwards
on average, then the average maximum score $\langle\maxscore\rangle$
is proportional to the logarithm of the sequence length as in the
logarithmic phase of gapless alignment. The reduced value of
$\langle\Sigma\rangle$ in the logarithmic phase makes it the regime of
choice for the purpose of homology detection.  Again, the distribution
of $\maxscore$ must be known for local alignments of random sequences
in order to characterize the statistical significance of local
alignment. There is no rigorous theory of this distribution in the
presence of gaps.  However, there is a lot of empirical evidence that
the distribution is again of the Gumbel
form~\cite{smit85,coll88,mott92,wate94a,wate94b,alts96}.  The values
of the parameters $\kappa$ and $\lambda$ are only known approximately
for a few cases close to the gapless limit~\cite{mott99,sieg99}. In
practice, they are determined empirically by time consuming
simulations. Below we will present an explicit calculation of the
parameter $\lambda$ for a simple scoring system.

\section{Significance Estimation using Global
Alignment}\label{sec_localtoglobal}

As a first step, we want to show that the parameter $\lambda$, which
describes the tail of the Gumbel distribution, can be derived solely
from studying the much simpler {\em global}
alignment~(\ref{eq_globrecurs}). Later, we will see that global
alignment is in certain cases equivalent to the asymmetric exclusion
process. We will derive an explicit formula for $\lambda$ by studying
the corresponding asymmetric exclusion process.

\begin{figure}[ht]
\begin{center}
\epsfig{figure=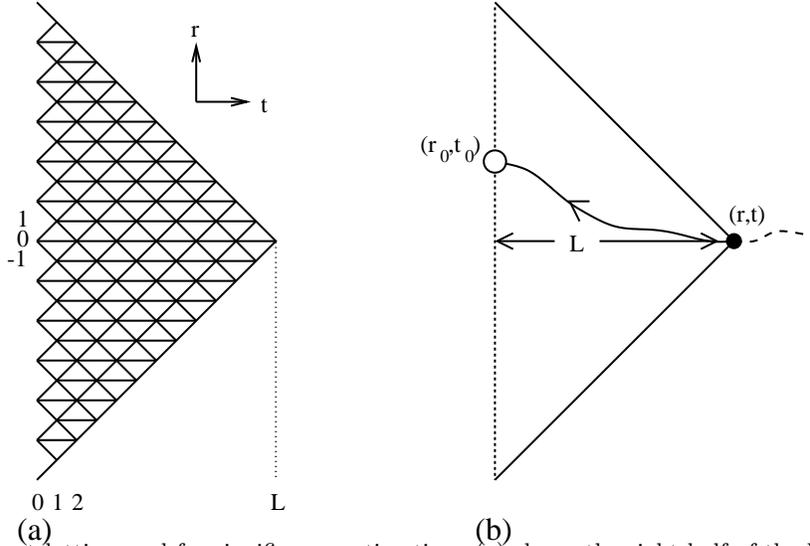,width=0.6\columnwidth}
\caption{Global alignment lattice used for significance estimation.
(a) shows the right half of the lattice from
Fig.~\protect\ref{fig_alpath}.  It can represent all possible paths of
length \protect$L$ which end at the point
\protect$(r,\timevar)=(0,L)$ and start at \protect$(r,0)$ for an
arbitrary \protect$r$. (b) shows such a path schematically. It
represents the ``rim'' of an island with its high score denoted by the
filled dot at the tip of the triangle. The open dot at
$(r_0,\timevar_0)$ represents the corresponding island initiation
event.}
\label{fig_triangle}
\end{center}
\end{figure}

Let us define the generating function
\begin{equation}\label{eq_defgf}
\gfgeneral(\lambda;L)\equiv\langle\exp[\lambda \score(r=0,L)]\rangle
\end{equation}
where the brackets $\langle\cdot\rangle$ denote the ensemble average
over all possible realizations of the disorder, i.e., over all choices
of random sequences $\sequenceone$ and $\sequencetwo$ and
$\score(0,L)$ is the {\em global} alignment score at the end of a
lattice of length $L$ as shown in Fig.~\ref{fig_triangle}(a). It can
be obtained from the recursion relation~(\ref{eq_globrecurs}) with the
initial condition
$\score(2k,\timevar=0)=\score(2k\!+\!1,\timevar\!=\!1)=0$. It will turn out
that the parameter $\lambda$ of the Gumbel distribution is obtained
from
\begin{equation}\label{eq_glambda}
\lim_{L\to\infty}\gfgeneral(\lambda;L)=1.
\end{equation}
Note, that this condition reduces simply to Eq.~(\ref{eq_lambdacond})
in the case of gapless alignment, since for infinite gap cost $\delta$,
we have
\begin{equation}
\langle\exp[\lambda \score(0,L)]\rangle=
\langle\exp[\lambda\sum_{k=1}^{L/2}\smatrix(0,2k-1)]\rangle=
\langle\exp[\lambda \smatrix]\rangle^{L/2}.
\end{equation}

\begin{figure}[ht]
\begin{center}
\epsfig{figure=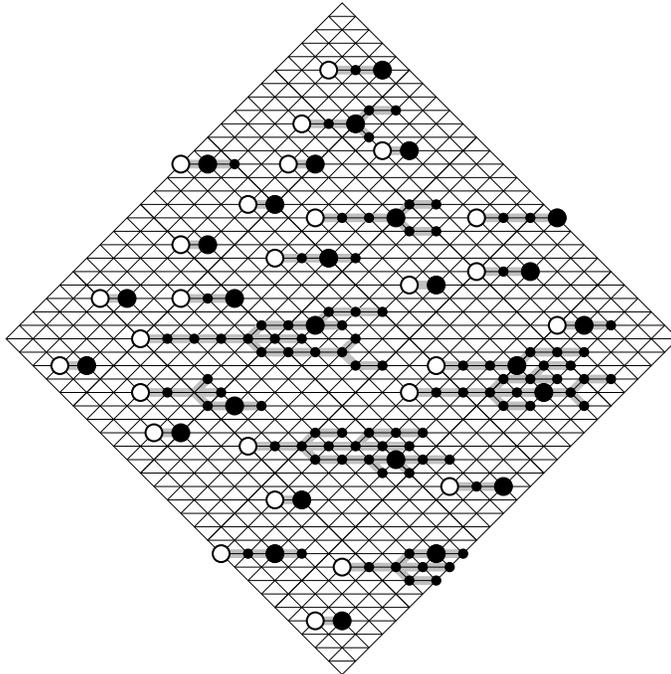,width=0.5\columnwidth}
\caption{Sketch of some islands on the local alignment lattice. The
lattice sites with a positive score are marked with dots. The bonds
which have been chosen in the maximization
process~(\protect\ref{eq_swrecursion}) are highlighted. Together they
are the restricted optimal path associated with each point with a
positive score. Each of these paths goes back to an island initiation
event which is marked by an open dot. The large filled dots mark the
positions of the highest scoring point on each island.}
\label{fig_islands}
\end{center}
\end{figure}

The key observation which leads to the result~(\ref{eq_defgf})
and~(\ref{eq_glambda}) is the fact that similar to the case of gapless
alignment discussed in the last section, the points on the alignment
lattice can be grouped together as {\em islands}~\cite{olse99}. By the
construction of the local alignment algorithm~(\ref{eq_swrecursion}),
many points on the alignment lattice have a score of zero in the
logarithmic alignment regime.  As for gapless alignment, a positive
score will be generated out of this ``sea'' of zeroes, if a good match
occurs by chance. This positive score can then imply further positive
scores via the recursion relation~(\ref{eq_swrecursion}). For every
point $(r,\timevar)$ on the lattice which has a positive score, we can
define a restricted optimal path $\widehat{r}^*_{r,\timevar}(\tau)$,
which is the highest scoring path out of all paths $\widehat{r}(\tau)$
with an end fixed at $\widehat{r}(\timevar)=r$; see the example in
Fig.~\ref{fig_alpath}. The path must start at some point
$(r_0,\timevar_0)$ where a positive score is created out of the zero
sea by a good match. An island is then defined to be the collection of
points $(r,\timevar)$ with positive score, i.e.,
$\localscore(r,\timevar)>0$, and whose restricted optimal path
$\widehat{r}^*_{r,\timevar}(\tau)$ originates at the same point
$(r_0,\timevar_0)$.  A sketch of these islands is shown in
Fig.~\ref{fig_islands}.  Each of these islands has a maximum score
which we denote by $\islandscore_k$ as we did in the gapless case. By
this definition, every lattice point with a positive score belongs to
exactly one island. Thus, the maximal score $\maxscore$ on the total
lattice is given by Eq.~(\ref{eq_sigmamax}).  Since large islands are
well separated by a sea of points with score zero, they are
statistically independent clusters on the alignment lattice. Thus,
their maximal scores $\islandscore_k$ are again independent
identically distributed random variables which yield a Gumbel
distribution of $\maxscore$ via Eq.~(\ref{eq_derivegumbel}). Our task
is thus to calculate the distribution of the island peak scores
$\islandscore_k$ in the presence of gaps.

This distribution of maximal island scores can be derived analogously
to the gapless case (App.~\ref{app_gllambda}.) While a single gapless
island is described by a random walk of some optimized length $L$, an
island with gaps corresponds to a {\em global gapped alignment} of some
optimized length $L$ as the one shown schematically in
Fig.~\ref{fig_triangle}(b). Using this replacement, the maximal island
distribution again has an asymptotically Poissonian
form~(\ref{eq_pislandpeak}) with the decay constant $\lambda$ given by
Eq.~(\ref{eq_glambda}). An approximate interpretation for the
result~(\ref{eq_glambda}) is the following: Due to the choice of
scoring parameters in the logarithmic phase of local alignment, the
average score $\langle\score(0,L)\rangle$ of global alignment with
the same choice of parameters decreases linearly with the length $L$
of the alignment. Thus, typical configurations of the disorder have a
strongly negative score $\score(0,L)$ and hardly contribute to
$\gfgeneral(\lambda;L)=\langle\exp[\lambda\score(0,L)]\rangle$. Only on
very rare occasions, $\score(0,L)$ is positive for large $L$ and
contributes significantly to $\gfgeneral(\lambda;L)$. The fact that
there is a choice of $\lambda$ with $\gfgeneral(\lambda;L)=1$ for large
$L$ implies that these configurations with positive $\score(0,L)$
are {\em exponentially rare}. It is thus necessary to weight these
configurations with the exponential factor
$\exp[\lambda\score(0,L)]$, and choose $\lambda$ to match the decay
constant of the probability of finding such rare events.

As already noted in the analogy between the directed polymer and
sequence alignment, the score corresponds to the (negative of the)
free energy.  Thus the quantity
$\gfgeneral(\lambda;L)=\langle\exp[\lambda \score(0,L)]\rangle$ can be
interpreted as the disorder-averaged (zero temperature) partition
function\footnote{However, $\gfgeneral(\lambda;L)$ should {\em not} be
interpreted as the partition function at temperature $\lambda^{-1}$.}
of $\lambda$ ``replicas'' of a directed polymer of length $L$. Note
that the replica number given by $\lambda$ need not be integer. In the
surface growth interpretation, $\gfgeneral(\lambda;L)$ is the
generating function for the space averaged surface height. While many
of the universal features of global and local sequence alignment
(e.g., its scaling behavior in the logarithmic phase and upon
approaching the phase transition line) can be understood merely from
the knowledge that sequence alignment belongs to the KPZ universality
class~\cite{hwa96,dras98a,hwa98,dras98b} or from the limit
$\gfgeneral(\lambda\to0;L)$, a solution of Eq.~(\ref{eq_glambda}) for
the non universal quantity $\lambda$ requires the knowledge of the
large $L$ behavior of the entire function $\gfgeneral(\lambda;L)$ and
hence a more detailed microscopic calculation for the given model.

\section{Global Alignment as an Asymmetric Exclusion
Process}\label{sec_maptoasep}

\subsection{A Simple Model of Sequence Alignment}

{From} now on we will focus on {\em global} alignment as described by
Eq.~(\ref{eq_globrecurs}), and use Eq.~(\ref{eq_glambda}) to infer the
value of the parameter $\lambda$ characterizing local alignment. In
order to simplify the presentation, we restrict ourselves here to a
very simple scoring system.  As will be discussed in latter sections,
our formulation as well as some of the results can be generalized to
the more complicated scoring systems.

We will study the scoring system in which the local similarity scores
$\smatrix_{\letterone,\lettertwo}$ can take on only two possible
values,
\begin{equation}
\smatrix_{\letterone,\lettertwo}=
\left\{\begin{array}{ll}1&\letterone=\lettertwo\\0&\letterone\not=\lettertwo
\end{array}\right..
\end{equation}
Moreover we will choose the gap cost to be $\delta=0$. With this
choice of the scoring parameters, the score $h$ has the additional
interpretation of being the length of the {\em longest common
subsequence} of the two sequences $\sequenceone$ and
$\sequencetwo$. This longest common subsequence problem has a long
history\footnote{The longest common subsequence model is in the limit
of an alphabet size equal to the sequence length also related to the
{\em longest increasing subsequence} model which has been of recent
interest in connection with surface growth processes~\cite{prae99}.}
as a toy model for sequence comparisons~\cite{chva75,danc94,monv99}.

Additionally, we will neglect correlations between the local scores
$\smatrix(r,\timevar)$, which arise from the fact that all $M\times
N$ local scores are generated by the $M+N$ randomly
drawn letters. Instead of taking these correlations into account, we
will assume that $\smatrix(r,\timevar)=\disorder(r,\timevar)$ with
independent random variables $\disorder(r,\timevar)$ given by
\begin{equation}\label{eq_etadist}
\disorder(r,t)=\left\{\begin{array}{ll}1&\mbox{with probab. $p$}\\
0&\mbox{with probab. $1-p$}\end{array}\right.
\end{equation}
with
\begin{equation}\label{eq_etaindep}
\Pr\{\forall_{r,\timevar}\,\disorder(r,\timevar)=\disorder_{r,\timevar}\}=
\prod_{r,\timevar}\Pr\{\disorder(r,\timevar)=\disorder_{r,\timevar}\}.
\end{equation}
To model sequences randomly drawn with equal probability from an
alphabet of size $c$, we take $p=1/c$. The
approximation~(\ref{eq_etaindep}) is known to change characteristic
quantities of sequence alignment only slightly~\cite{dras98a}.  We
will confirm numerically at the end of this paper, that this also
holds for the values of $\lambda$ which we are mainly interested in
here. For our choices of parameters, the global alignment
algorithm~(\ref{eq_globrecurs}) reads
\begin{equation}\label{eq_lcsrecurs}
\score(r,\timevar+1)=
\max\{\score(r,\timevar-1)+\eta(r,\timevar),
\score(r+1,\timevar),\score(r-1,\timevar)\}.
\end{equation}

\subsection{Choice of the alignment lattice geometry}

\begin{figure}[ht]
\begin{center}
\epsfig{figure=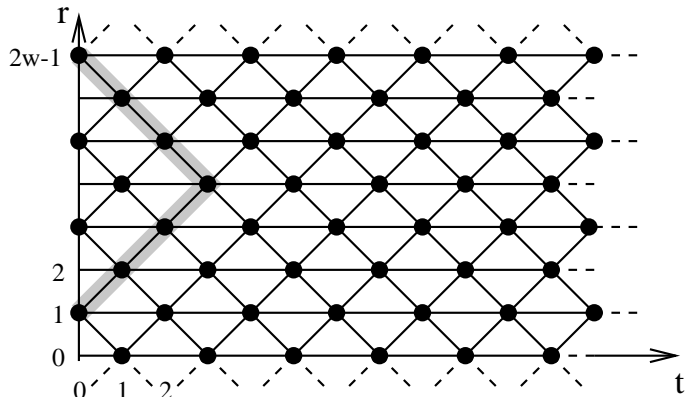,width=0.5\columnwidth}
\caption{Rectangular alignment lattice of width \protect$2\width$ with
periodic boundary conditions in the spatial (vertical) direction. We
use this lattice instead of the triangular lattice shown in
Fig.~\protect\ref{fig_triangle}(a) in order to simplify the handling of
finite-size effects. As indicated by the thick gray lines, the score at a
point with \protect$\timevar\le\width$ as the one at the tip of the
triangle is identical with the corresponding score calculated on a
triangular lattice as the one shown in
Fig.~\protect\ref{fig_triangle}(a).}
\label{fig_rect}
\end{center}
\end{figure}

In order to handle finite size effects better, we will use a
rectangular geometry (Fig.~\ref{fig_rect}) for the alignment lattice,
instead of the triangular geometry shown in
Fig.~\ref{fig_triangle}(a).  We will further apply periodic boundary
condition to the top and bottom edges of the lattice, i.e.,
$\score(0,\timevar)=\score(2\width,\timevar)$ for a rectangular
lattice of width $2\width$, and will start on the left edge with the
initial conditions $\score(2k+1,\timevar=0)=\score(2k,\timevar=1)=0$.
Note that despite the different lattice geometries, the score
$\score(r,\timevar)$ for all points with $\timevar\le\width$ on the
rectangular lattice will be {\em identical} to the score at the same
$(r,\timevar)$ coordinate on the triangular lattice\footnote{Since
directed polymers in a random medium are known to have a wandering
exponent $\zeta=2/3$ this actually still holds for
$\timevar<\width^{3/2}$.}; see Fig.~\ref{fig_rect}.

\subsection{The dynamics of sequence alignment as an asymmetric exclusion
process}

\begin{figure}[htbp]
%\narrowtext
\begin{center}
\epsfig{figure=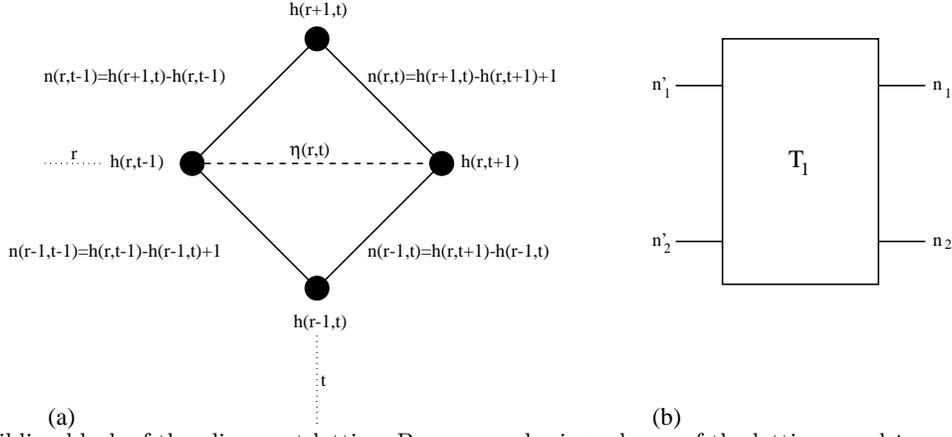,width=0.7\columnwidth}
%FIGSPACE:60mm
\caption{One building block of the alignment lattice. By our numbering
scheme of the lattice $r$ and $\timevar$ are either both even or both
odd. (a) shows the scores at the lattice points and the bond variables
\protect$\occupation(r,\timevar)$. (b) shows this building block as a
``device'', which takes two incoming bond variables
\protect$n_1^\prime$ and \protect$n_2^\prime$ and transforms them with
the help of the transfer matrix \protect${\sl T}_0$ into the new bond
variables \protect$n_1$ and \protect$n_2$.}
\label{fig_element}
\end{center}
\end{figure}

In this section we will perform a change of variables on the sequence
alignment algorithm~(\ref{eq_lcsrecurs}) for the rectangular lattice
shown in Fig.~\ref{fig_rect}. We will find that the resulting problem
is equivalent to an asymmetric exclusion process on a one-dimensional
lattice of width $2\width$.  As a guidance towards the choice of
suitable variables, we take the knowledge from the (continuous) KPZ
equation that the gradient of the surface height is an especially
simple quantity. At a fixed time, the gradients at different positions
become uncorrelated and Gaussian
distributed~\cite{kard86,krug91b}. Thus, we will look at their
discrete analogs in the alignment problem. They are the score
differences between neighboring lattice points and thus located on the
diagonal bonds of the lattice. We will parameterize these score
differences by the bond variables $\occupation(r,\timevar)$.  They
will later turn out to be the occupation numbers of the sites of an
asymmetric exclusion process. With the choice of coordinates as
illustrated in Fig.~\ref{fig_element}(a), we define them to
be\footnote{Note, that the $\occupation(r,\timevar)$ are not literally
score differences but suitably chosen parameterizations of these score
differences.  This complication is necessary in order to enable the
interpretation as the particle occupation numbers in the asymmetric
exclusion process.}
\begin{equation}\label{eq_occupdiff}
\occupation(r,\timevar)\equiv\left\{
\begin{array}{ll}\score(r+1,\timevar)-\score(r,\timevar+1)+1&
\mbox{for $r+\timevar$ even}\\[5pt]
\score(r+1,\timevar+1)-\score(r,\timevar)&
\mbox{for $r+\timevar$ odd}\end{array}\right.
\end{equation}

As explained in detail in App.~\ref{app_vtransform}, rewriting the time
evolution equation~(\ref{eq_lcsrecurs}) in terms of the variables
$\occupation(r,\timevar)$ leads to a time evolution equation of
$\occupation(r,\timevar)$ alone, without any reference to the absolute
scores $\score(r,\timevar)$. Moreover, this time evolution equation
implies that the score differences take only the values
$\occupation(r,\timevar)\in\{0,1\}$. By the structure of the alignment
lattice as a composition of elements as the one shown in
Fig.~\ref{fig_element}(a), the resulting time evolution for the
$\occupation(r,\timevar)$ transforms a pair
$(\occupation(r-1,\timevar-1),\occupation(r,\timevar-1))\in\{|00\rangle,
|01\rangle,|10\rangle,|11\rangle\}$ into the new pair
$(\occupation(r-1,\timevar),\occupation(r,\timevar))\in\{|00\rangle,
|01\rangle,|10\rangle,|11\rangle\}$ independently from all the other
$\occupation(r^\prime,\timevar-1)$. This transformation only depends
on the single random variable $\disorder(r,\timevar)$ and can be
expressed by the transfer matrix
\begin{equation}\label{eq_transfermatrix}
{\sf T}_1(0)\equiv\left(\begin{array}{cccc}1&0&0&0\\0&1&1-p&0\\0&0&p&0\\0&0&0&1
\end{array}\right)
\end{equation}
in the basis $|00\rangle$, $|01\rangle$, $|10\rangle$, $|11\rangle$.
We can thus interpret the action of the lattice element shown in
Fig.~\ref{fig_element}(a) as a ``device'' like the one shown in
Fig.~\ref{fig_element}(b) which takes a pair of variables
$(\occupation_1^\prime,\occupation_2^\prime)$ as its inputs, applies
the transfer matrix ${\sf T}_1(0)$, and generates a new pair of
variables $(\occupation_1,\occupation_2)$ as its outputs.

\begin{figure}[htbp]
%\narrowtext
\begin{center}
\epsfig{figure=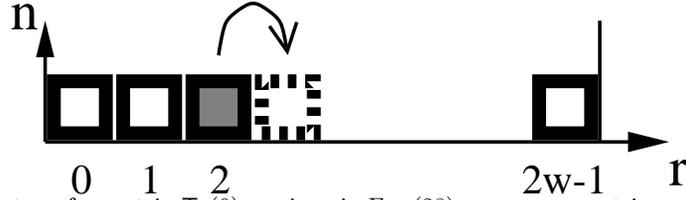,width=0.5\columnwidth}
%FIGSPACE:60mm
\caption{Interpretation of the transfer matrix \protect${\sf T}_1(0)$ as given
in Eq.~(\protect\ref{eq_transfermatrix}) as
an asymmetric exclusion process. A configuration of the
local score differences is represented by particles on a
one-dimensional lattice of width \protect$2\width$.
At an odd time step for each even site \protect$r-1$
a particle hop is attempted with probability \protect$1-p$. In our
example, the particle at site \protect$0$ cannot hop, since site \protect$1$
is already occupied. The particle on site \protect$2$ can hop to site
\protect$3$ as indicated by the dashed square.}
\label{fig_singleparticles}
\end{center}
\end{figure}

We recognize the action of the transfer matrix ${\sf T}_1(0)$ as the
elementary time step of an asymmetric exclusion process, if we
interpret the $\occupation(r,\timevar)$ as particle occupation numbers
on a one-dimensional lattice of $2\width$ sites with periodic boundary
conditions as the one shown in Fig.~\ref{fig_singleparticles}.  Each
of these sites can either be empty or occupied by a single particle.
In each time step for each pair of neighboring sites, a particle hops
to the right with some probability $1-p$, if the site to its right is
empty according to the non vanishing entry $|10\rangle\to|01\rangle$
of the transfer matrix ${\sf T}_1(0)$. If there is no particle or if
the site on the right is already occupied the configuration remains
unchanged.

\begin{figure}[htbp]
%\narrowtext
\begin{center}
\epsfig{figure=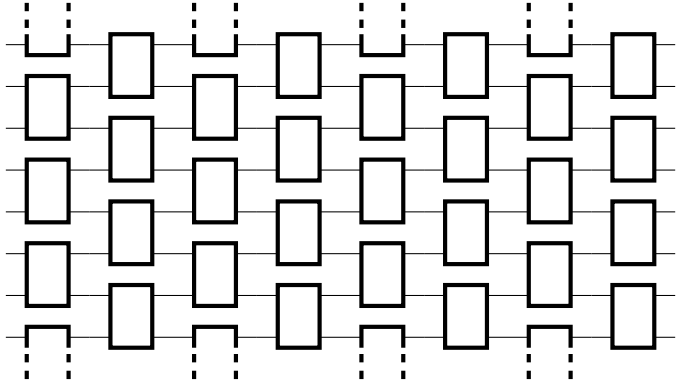,width=0.5\columnwidth}
%FIGSPACE:60mm
\caption{Schematic representation of the alignment lattice of
Fig.~\protect\ref{fig_rect} as an ``electric circuit''.  The boxes
represent elements of the type shown in
Fig.~\protect\ref{fig_element}(b). They take two particle occupation
numbers as their ``inputs'' and generate two new particle occupation
numbers as their ``outputs''. Their interconnection into a layered
structure as shown here with a shifted pairing scheme in every time
step leads to the non-trivial behavior of sequence alignment.}
\label{fig_network}
\end{center}
\end{figure}

In terms of the elementary devices shown in Fig.~\ref{fig_element}(b)
the lattice structure of Fig.~\ref{fig_rect} can be depicted
schematically as shown in Fig.~\ref{fig_network}. Thus, the process of
hopping a particle to the right is attempted for each even numbered
site at odd time steps and for each odd numbered site at even time
steps. This hopping dynamics is exactly the asymmetric exclusion
process with sublattice-parallel updating with periodic boundary
conditions\footnote{If we had chosen the ``hard wall'' boundary
conditions $\score(-1,t)=\score(2\width,t)=\infty$ instead of the
periodic boundary conditions $\score(2\width,t)=\score(0,t)$ for the
score, we would have arrived at the asymmetric exclusion process with
sublattice-parallel updating and {\em open} boundary conditions at a
feeding and extinction rate of $\alpha=\beta=1-p$ at the two ends of
the lattice with $2\width-1$ sites
respectively.}~\cite{kand90,raje98}.

In reducing the dynamics from a dynamics of scores into a dynamics of
the occupation numbers $\occupation(r,\timevar)$, one has to pay
attention to the boundary conditions. Periodic boundary conditions for
the $\occupation(r,\timevar)$ do not automatically lead to meaningful
periodic boundary conditions for the scores $\score(r,\timevar)$.  We
thus have to impose the additional constraint that the total sum of
the local score differences across the whole lattice vanishes. In
terms of our bond variables $\occupation(r,\timevar)$ this translates
into the condition
\begin{equation}\label{eq_initialconstraint}
\frac{1}{2\width}\sum_{r=0}^{2\width-1}\occupation(r,\timevar)=\frac{1}{2},
\end{equation}
i.e., the system of hopping particles is at half filling. Since the
number of particles is conserved under the dynamics described by the
transfer matrix ${\sf T}_1(0)$, the condition~(\ref{eq_initialconstraint})
is guaranteed to hold if we choose the initial conditions
$\sum_{r=0}^{2\width-1}\occupation(r,\timevar=0)/2\width=1/2$.
Particle densities different from one half would correspond to a
tilted ``score profile'' $\score(r,\timevar)$ at each fixed time
$\timevar$.

\section{The Generating Function}\label{sec_evalgenfunc}

\subsection{Expressing the generating function in terms of the hopping
process}

We now want to apply the mapping between sequence alignment and the
asymmetric exclusion process to the practical problem of assessing
alignment significance. As noted in Sec.~\ref{sec_localtoglobal}, this
amounts to calculating the generating function
\begin{equation}
\gfgeneral_0(\lambda;N)\equiv\langle\exp[\lambda\score(0,N)]\rangle_0,
\end{equation}
where $\langle\ldots\rangle_0$ denoted the average over the ensemble
of uncorrelated disorder defined by Eqs.~(\ref{eq_etadist})
and~(\ref{eq_etaindep}).  Thus, we first need to express the total
score $\score(0,N)$ in terms of the occupation numbers
$\occupation(r,\timevar)$.  As explained in more detail in
App.~\ref{app_vtransform}, $\score(0,\timevar)$ is on average
incremented by $1/2\width$ every time the transfer matrix ${\sf
T}_1(0)$ is applied except for the transition
$|01\rangle\to|10\rangle$. Thus, $\gfgeneral_0(\lambda;N)$ can be
expressed as
\begin{equation}\label{eq_explambdas}
\gfgeneral_0(\lambda;N)=
\exp[\lambda N/2]\langle\exp[-\lambda\totalhops]\rangle_0
\end{equation}
in terms of the total number of particle hops per lattice site
\begin{equation}
\totalhops\equiv\frac{1}{2\width}\sum_{l=1}^{N/2}\sum_{k=0}^{\width-1}
(\hopnumber(2k+1,2l-1)+\hopnumber(2k,2l)),
\end{equation}
where $\hopnumber(r,\timevar)\in\{0,1\}$ is the number of particle
hops at lattice site $(r,\timevar)$.  We thus need to
determine the generating function
\begin{equation}\label{eq_genfuncsbyeachother}
\genfunc(\lambda;\width,N)\equiv\langle\exp[-\lambda\totalhops]\rangle_0
\end{equation}
for the asymmetric exclusion process. Note, that this is different
from the generating function of the local current
$\hopnumber(r,\timevar)$: since $\totalhops/N$ is the {\em time and
space averaged} current, $\genfunc$ contains information on spatial
and temporal {\em correlations} in the number of hopping particles
which the generating function for the local current does not.

\subsection{The Generating function as an eigenvalue
problem}\label{sec_eigenvalue}

Now we will reformulate the calculation of the generating function
$\genfunc(\lambda;\width,N)$ for the asymmetric exclusion process as
an eigenvalue problem. As already mentioned,
$\exp[-\lambda\totalhops]$ is a product of factors
$\exp[-\lambda/2\width]$ for every particle that hops. Since the
dynamics of the hopping process is described by the transfer matrix
${\sf T}_1(0)$ defined in Eq.~(\ref{eq_transfermatrix}), we can
calculate $\genfunc(\lambda;\width,N)$ by associating a weight
$\exp[-\lambda/2\width]$ to the element of the transfer matrix ${\sf
T}_1(0)$ which corresponds to a hop. This can be derived more formally
from a dynamics path integral representation of
$\gfgeneral_0(\lambda;N)$ as detailed in App.~\ref{app_pathintegral}. We
get the modified local transfer matrix
\begin{equation}\label{eq_deftoflambda}
{\sf T}_1\left(\frac{\lambda}{\width}\right)\equiv\left(\begin{array}{cccc}
1&0&0&0\\0&1&(1-p)e^{-\frac{\lambda}{2\width}}&0\\0&0&p&0\\0&0&0&1
\end{array}\right)
\end{equation}
in the basis $|00\rangle,|01\rangle,|10\rangle,|11\rangle$ of a pair
of neighboring lattice sites.

Next, we need to take into account the special lattice structure of
Fig.~\ref{fig_network}. We note that at every even time step the
lattice is decomposed into $\width$ of the building blocks described
by ${\sf T}_1$. Thus, a single time step of the total system
at even time is described by the matrix
\begin{equation}
{\sf T}_{\mathrm even}\equiv
{\sf T}_\width(\lambda)\equiv
\bigotimes_{k=1}^\width {\sf T}_1\left(\frac{\lambda}{\width}\right).
\end{equation}
At odd times the dynamics is the same, but the pairing of neighboring
sites is shifted. To generate the time evolution at odd time steps, we
can thus shift all particles to the right, apply the dynamics of even time
steps and then shift all particles back to the left. Let
${\sf C}$ be the translation operator such that
\begin{equation}
{\sf C}|\occupation_0\occupation_1\ldots\occupation_{2\width-1}\rangle
\equiv|\occupation_1\ldots\occupation_{2\width-1}\occupation_0\rangle
\end{equation}
which shifts all particles by one site to the left taking into account
the periodic boundary conditions.  With this definition we can write
${\sf T}_{\mathrm odd}={\sf C}{\sf T}_\width(\lambda){\sf C}^{-1}$.

The sublattice-parallel updating procedure (i.e., the structure of the
lattice as depicted by Fig.~\ref{fig_network}) finally leads to
\begin{equation}\label{eq_genasmatrixproduct}
\genfunc(\lambda;\width,N)=
\langle\psi_1|({\sf T}_{\mathrm even}
{\sf T}_{\mathrm odd})^{N/2}|\psi_0\rangle=
\langle\psi_1|({\sf T}_\width(\lambda){\sf C}{\sf T}_\width(\lambda)
{\sf C}^{-1})^{N/2}|\psi_0\rangle
\end{equation}
where $|\psi_0\rangle$ is a $4^{\width}$ dimensional state vector
representing the initial conditions, and $\langle\psi_1|$ is the
$4^{\width}$ dimensional vector whose entries are all $1$, used here
to denote a summation over all possible final configurations.  In the
limit of large $N\gg\width$, this obviously becomes
\begin{equation}\label{eq_rhodef}
\genfunc(\lambda;\width,N)=\eigenval_\width^{N}(\lambda)
\end{equation}
where $\eigenval_\width^2(\lambda)$ is the eigenvalue of ${\sf
T}_\width(\lambda){\sf C}{\sf T}_\width(\lambda){\sf C}^{-1}$ with the
largest real part.  Since this matrix has no negative entries and is
irreducible for non-pathological choices of the scoring matrix (while
restricted to the physical sector of half filling), the largest
eigenvalue of this matrix is guaranteed by the Perron Frobenius
theorem to be non degenerate and real, and its eigenvector can be
chosen without negative entries.  When $\lambda=0$, we have
$\eigenval(0)=1$ and its eigenvector is the stationary distribution of
the asymmetric exclusion process, which is a simple tensor product of
independent occupation numbers. This is no longer the case for
$\lambda\not=0$.

\subsection{Calculating the largest eigenvalue}

For a finite $\width$, it is in principle possible to solve for the
largest eigenvalue of the $4^\width$ dimensional matrix ${\sf
T}_\width(\lambda){\sf C}{\sf T}_\width(\lambda){\sf C}^{-1}$ by
directly diagonalizing the matrix.  It is convenient to reduce the
size of this matrix by exploiting some symmetries. Since the lattice
is translationally invariant with respect to shifts in $r$ by $2$, we
expect the same symmetry of the largest eigenvalue of ${\sf
T}_\width(\lambda){\sf C}{\sf T}_\width(\lambda){\sf C}^{-1}$. Thus,
for the purpose of computing the largest eigenvalue we can restrict
ourselves to the subspace ${\cal C}$ of translationally invariant
vectors
\begin{equation}
{\cal C}\equiv\Big\{|\psi\rangle\Big|C^2|\psi\rangle=|\psi\rangle\Big\}.
\end{equation}
This corresponds to a discrete Fourier transform of the matrix ${\sf
T}_\width(\lambda){\sf C}{\sf T}_\width(\lambda){\sf C}^{-1}$ and
choosing the $k=0$ component. On ${\cal C}$, we have ${\sf
C}^{-1}={\sf C}$ by definition. Thus, it is enough to look for the
largest eigenvalue $\eigenval_\width(\lambda)$ of the matrix ${\sf
T}_\width(\lambda){\sf C}$ restricted to ${\cal C}$.  A further
restriction which helps reducing the size of the matrix is the mirror
symmetry of the lattice which has to be respected by the eigenvector
as well. Additionally, ${\sf T}_\width(\lambda){\sf C}$ has to be
restricted onto the physical subspace of half filling.

After applying these simplifications, the largest eigenvalue can be
calculated for small widths $\width$ using computer algebra.  Although
the matrix ${\sf T}_\width(\lambda){\sf C}$ explicitly contains the
quantity $\exp[-\lambda/2\width]$, it turns out that the
characteristic polynomial depends only on $\exp[-\lambda/2]$. This is
a consequence of the translational invariance of the
lattice\footnote{Instead of looking at the average score
$\overline{\score}(N)=\frac{1}{2\width}\sum_r\score(r,N)$ as we do in
the derivation of Eq.~(\ref{eq_genasmatrixproduct}) in
App.~\ref{app_pathintegral}, we could also have chosen a specific
position, say $r=0$ and $r=1$, and monitored the behavior of the score
$\widetilde{\score}(N)\equiv\frac{1}{2}[\score(1,N)+\score(0,N-1)]$.
Since the differences between scores at the same time are bounded,
these two quantities must have the same generating function for large
$N$. The transfer matrix which calculates the generating function for
$\widetilde{\score}(N)$ is $\widetilde{{\sf T}}(\lambda)\equiv {\sf
T}_1(\lambda)\otimes \bigotimes_{k=2}^{\width}{\sf T}_1(0)$ instead of
${\sf T}_\width(\lambda)$. It has the technical disadvantage that it
breaks the translational invariance, but it explicitly depends only on
$\exp[-\lambda/2]$ instead of $\exp[-\lambda/2\width]$.}.  In order to
reveal the underlying structure of the largest eigenvalues for
different $\width$, it is very useful to {\em expand} the resulting
largest eigenvalues $\eigenval_\width(\lambda)$ in powers of this
quantity $e^{-\lambda/2}$. We get
\begin{eqnarray*}
\width=1:\quad \eigenval_1(\lambda)&=&\sqrt{p}+O(e^{-\frac{\lambda}{2}})\\
\width=2:\quad \eigenval_2(\lambda)&=&\sqrt{p}-(p-1)e^{-\frac{\lambda}{2}}
+O((e^{-\frac{\lambda}{2}})^2)\\
\width=3:\quad \eigenval_3(\lambda)&=&\sqrt{p}-(p-1)e^{-\frac{\lambda}{2}}
+(p-1)\sqrt{p}(e^{-\frac{\lambda}{2}})^2+O((e^{-\frac{\lambda}{2}})^3)\\
\width=4:\quad \eigenval_4(\lambda)&=&\sqrt{p}-(p-1)e^{-\frac{\lambda}{2}}
+(p-1)\sqrt{p}(e^{-\frac{\lambda}{2}})^2
-(p-1)\sqrt{p}^2(e^{-\frac{\lambda}{2}})^3+
O((e^{-\frac{\lambda}{2}})^4),
\end{eqnarray*}
where the $O((e^{-\lambda/2})^k)$ terms denote terms of the given
order with prefactors which are different for different $\width$. We
can see that the coefficients up to order
$(e^{-\lambda/2})^{\width-1}$ remain unchanged upon increasing
$\width$ and they constitute the beginning of a simple geometric
series. Assuming that this pattern holds for arbitrary orders, we can
resum the series for any {\em fixed} $\lambda>0$ and get
\begin{equation}
\eigenval(\lambda)\equiv\lim_{\width\to\infty}\eigenval_\width(\lambda)=
\frac{\sqrt{p}+e^{-\frac{\lambda}{2}}}{1+\sqrt{p}e^{-\frac{\lambda}{2}}}.
\end{equation}
Combined with Eqs.~(\ref{eq_explambdas}),
(\ref{eq_genfuncsbyeachother}), and~(\ref{eq_rhodef}) this yields the
generating function
\begin{equation}\label{eq_sfunclcs}
\gfgeneral_0(\lambda;N)=\exp[\lambda N/2]\eigenval^{N}(\lambda)=
\left(\exp[\lambda/2]\eigenval(\lambda)\right)^{N}=
\left(\frac{1+\sqrt{p}\exp[\frac{\lambda}{2}]}%
{1+\sqrt{p}\exp[-\frac{\lambda}{2}]}\exp[-\frac{\lambda}{2}]\right)^{N}
\end{equation}
in the limit of large $N$. 

A related generating function has also recently~\cite{derr98b} been
calculated in the simpler case of a discrete space and continuous time
asymmetric exclusion process. In this case, the full dependence on the
finite width $\width$ can be computed. While Derrida {\it et
al.}~\cite{derr98b} find that the generating function takes a {\em
universal} form in the limit $\width\to\infty$ with
$\lambda\width^{1/2}$ kept constant, the problem of assessing
statistical significance of sequence alignments calls for the limit
$\width\to\infty$ at a fixed $\lambda>0$. This limit goes {\em beyond}
the universal regime. For the asymmetric exclusion process with
sublattice-parallel updating it is given by our
Eq.~(\ref{eq_sfunclcs})

Eq.~(\ref{eq_sfunclcs}) can be generalized~\cite{bund99b} to the
match-mismatch scoring system given in Eq.~(\ref{eq_idmatrix}) with a
gap cost $\delta=\mu/2$ for an arbitrary value of $\mu$. If we denote
the score in this scoring system by $\score^\prime(r,\timevar)$ it is
connected to the score $\score(r,\timevar)$ of the scoring system with
$\mu=\delta=0$ by the simple global rescaling and shifting
\begin{equation}
\score^\prime(r,\timevar)=(1+\mu)\score(r,\timevar)-\frac{\mu}{2}\timevar.
\end{equation}
Thus the corresponding generating function is given by
\begin{equation}
\gfgeneral(\lambda,\mu;N)\equiv\langle e^{\lambda\score^\prime(0,N)}\rangle
=e^{-\mu N}\langle e^{\lambda(1+\mu)\score(0,N)}\rangle.
\end{equation}
If we again neglect correlations and use uncorrelated random variables
\begin{equation}
\eta(r,\timevar)=\left\{\begin{array}{ll}1&\mbox{with probab. $p$}\\
-\mu&\mbox{with probab. $1-p$}\end{array}\right.
\end{equation}
the same rescaling and shifting leads to
\begin{equation}\label{eq_gfresult}
\gfgeneral_0(\lambda,\mu;N)\equiv
\langle e^{\lambda\score^\prime(0,N)}\rangle_0=
\left(\frac{1+\sqrt{p}\exp[\frac{\lambda}{2}(1+\mu)]}%
{1+\sqrt{p}\exp[-\frac{\lambda}{2}(1+\mu)]}\exp[-\frac{\lambda}{2}\mu]
\right)^{N}.
\end{equation}

\section{Implications on directed polymers and sequence
alignment}\label{sec_implications}

Now, we will study the consequences of our main result
Eq.~(\ref{eq_gfresult}). First, we will discuss the general properties
of the generating function and its implications on the physics of
directed polymers in a random medium. Then, we will come back to our
original question of the assessment of sequence alignment
significance.  We find, that Eq.~(\ref{eq_gfresult}) is an explicit
expression for the significance assessment parameter $\lambda$. It
reproduces known limiting cases and we will demonstrate that our
result agrees well with numerical simulations.

\subsection{Properties of the generating function}

The most notable property of the generating function of the connected
moments of the average score (or average height)
\begin{equation}
\log\langle\exp[\lambda\score^\prime(0,N)]\rangle_0=
\log\gfgeneral_0(\lambda,\mu;N)
\end{equation}
is that it is an {\em odd} function of $\lambda$. The first two terms
of its expansion are
\begin{equation}\label{eq_freeenergy}
\frac{\log\gfgeneral_0(\lambda,\mu;N)}{N}=
v(\mu)\lambda+\frac{1}{6}b(\mu)\lambda^3+O(\lambda^5).
\end{equation}
where
\begin{equation}\label{eq_velocity}
v(\mu)=\left.\frac{\mathrm d}{{\mathrm d}\lambda}\right|_{\lambda=0}
[\gfgeneral_0(\lambda,\mu;N)]^{\frac{1}{N}}=
-\frac{\mu}{2}+(1+\mu)\frac{\sqrt{p}}{1+\sqrt{p}}
\end{equation}
and $b(\mu)=\left(\frac{1+\mu}{1+\sqrt{p}}\right)^3
\frac{(1-\sqrt{p})\sqrt{p}}{4}>0$.  As already mentioned, we can regard
the generating function $\gfgeneral_0(\lambda,\mu;N)$ as the ensemble
averaged partition function of $\lambda$ replicas of a directed
polymer in a random medium. In this sense, Eq.~(\ref{eq_freeenergy})
is the free energy per length of this $\lambda$ replica system. It has
the same form (with a vanishing quadratic term) as the result of an
earlier explicit replica calculation in continuous time and continuous
space~\cite{kard85}.  However, our analysis is directly on the
discrete model and is not plagued by the difficulty of taking the
continuum limit in~\cite{kard85}. Moreover, we do not have to rely on
a questionable analytic continuation in the replica number since our
calculation is valid for an arbitrary $\lambda>0$ from the
very beginning.

The vanishing of the second order term in $\lambda$ will not even be
affected by the universal contributions to our result for small
$\lambda$ which have been found in~\cite{derr98b} using the explicit
dependence on the width $\width$, since its second order coefficient
vanishes as $\width^{-1/2}$ in the limit of large width.  The
consequence of this vanishing second order term in $\lambda$ is that
the second connected moment of the average height, i.e., the height
fluctuations, scales sublinear in $N$.  Instead the {\em third} moment
of the height fluctuations scales linearly with $N$. This is a
signature of the presence of the anomalous $N^{1/3}$ fluctuations of
the average surface height characteristic for the KPZ universality
class.

\subsection{Statistical significance and the log-linear transition}

According to Eqs.~(\ref{eq_glambda}) and~(\ref{eq_gfresult}) the
parameter $\lambda$ which characterizes the statistical significance
of local alignments with the match-mismatch scoring scheme
Eq.~(\ref{eq_idmatrix}) and gap cost $\delta=\mu/2$ is given by the
unique positive solution of the equation
\begin{equation}\label{eq_lambdaresult}
\frac{1+\sqrt{p}\exp[\frac{\lambda}{2}(1+\mu)]}%
{1+\sqrt{p}\exp[-\frac{\lambda}{2}(1+\mu)]}\exp[-\frac{\lambda}{2}\mu]=1.
\end{equation}
In the limit of large $\mu$, the solution of
Eq.~(\ref{eq_lambdaresult}) converges to $\lambda=-\log{p}$. This is
the value which we expect since this limit corresponds to the case of
gapless alignment (recall that $\delta=\mu/2$ here), 
and $\lambda=-\log{p}$ is the solution of the large
$\mu$ limit of Eq.~(\ref{eq_lambdacond}). If the gap cost is
decreased, $\lambda$ is reduced, too. At some critical value of $\mu$
there will not be any positive solution of Eq.~(\ref{eq_lambdaresult})
any more, i.e., islands of all sizes are equally probable. This indicates
a phase transition between the logarithmic and the linear alignment
phase. The approach of this phase transition is especially interesting.

Close to the phase transition, we can use the expansion~(\ref{eq_freeenergy})
and rewrite Eq.~(\ref{eq_lambdaresult}) as
\begin{equation}\label{eq_logcondition}
v(\mu)\lambda+\frac{1}{6}b(\mu)\lambda^3+O(\lambda^5)=0.
\end{equation}
{From} this expansion the origin of the phase transition is very
clear: If $v(\mu)>0$, the right hand side of
Eq.~(\ref{eq_logcondition}) is a monotonously increasing function of
$\lambda$. Thus, $\lambda=0$ is the only solution of
Eq.~(\ref{eq_logcondition}). This corresponds to a flat distribution
of island sizes, i.e., the linear alignment phase.  If $v(\mu<0)$, the
shape of the right hand side of Eq.~(\ref{eq_logcondition}) changes
and there are three roots, one of which is the positive solution
\begin{equation}\label{eq_lambdaapprox}
\lambda\approx\left(-6\frac{v(\mu)}{b(\mu)}\right)^{1/2}.
\end{equation}
This indicates that we are in the logarithmic alignment phase. Thus,
the phase transition occurs at the critical mismatch cost
$\mu_{\mathrm c}$ which is defined by the condition
\begin{equation}
v(\mu_{\mathrm c})=0.
\end{equation}
Using the explicit form~(\ref{eq_velocity}) of $v(\mu)$, we get the
critical mismatch cost
\begin{equation}
\mu_{\mathrm c}=\frac{2\sqrt{p}}{1-\sqrt{p}}.
\end{equation}
This reproduces the already known result~\cite{bund99} for the phase
transition point of this model. As the mismatch cost $\mu$ approaches
this critical value from above, $\lambda$ vanishes as
\begin{equation}
\lambda\approx\left(\frac{6(1-\sqrt{p})^3}{\sqrt{p}(1+\sqrt{p})}\right)^{1/2}
(\mu-\mu_{\mathrm c})^{1/2}.
\end{equation}
In the case of finite width $\width$, the above expression is valid down to
$\lambda\sim\width^{-1}$.  This confirms the characteristic universal
power law $|\mu-\mu_{\mathrm c}|^{1/2}$ 
proposed previously~\cite{dras98b} by scaling arguments.

\subsection{Numerical Verification}

\begin{figure}[htbp]
%\narrowtext
\begin{center}
\epsfig{figure=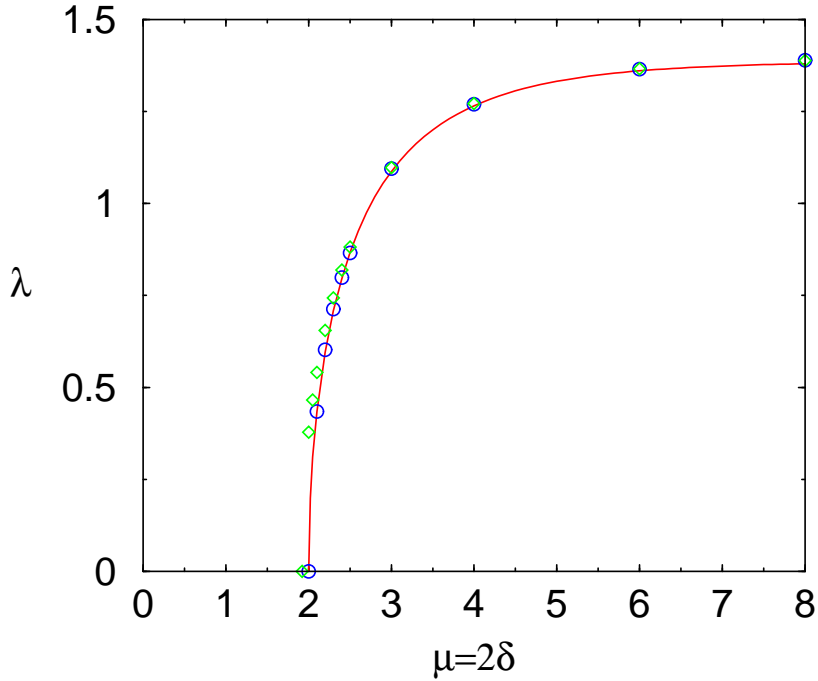,angle=270,width=0.6\columnwidth}
%FIGSPACE:60mm
\caption{Dependence of the significance parameter \protect$\lambda$ on
the scoring parameter \protect$\mu$. The circles represent the
numerically obtained values of \protect$\lambda$ for uncorrelated
local disorder~(\ref{eq_etaindep}) with match probability
\protect$p=1/4$ for which Eq.~(\protect\ref{eq_lambdaresult}) (the
solid line) has been derived.  They agree well with the analytical
result. The diamonds correspond to local disorder generated by
comparing two randomly chosen sequences over an alphabet of size
\protect$c=4$. The values of \protect$\lambda$ obtained from the two
ensembles differ from each other only very close to the phase
transition point \protect$\mu_{\mathrm c}=2$.}
\label{fig_lambdanumerics}
\end{center}
\end{figure}

In order to test the approximation of uncorrelated local
disorder~(\ref{eq_etaindep}) and the heuristic elements of the
derivation of Eq.~(\ref{eq_lambdaresult}), we performed extensive
numerical simulations to corrobate our result. We used the DNA
alphabet of size $c=4$ with identical frequencies for all four
letters, i.e., $p=1/4$. For different choices of the mismatch cost $\mu$
with corresponding gap cost $\delta=\mu/2$, we used the island
method~\cite{olse99} to find the values of $\lambda$ as a function of
$\mu$ numerically.  For each value of $\delta$ several billion
islands have been generated using sequences of $N=25,000$ in order to
achieve relative errors of approximately $1\%$. We used completely
uncorrelated local scores chosen as
\begin{equation}
\smatrix(r,\timevar)=\left\{\begin{array}{ll}1&\mbox{with probab. $p$}\\
-\mu&\mbox{with probab. $1-p$}\end{array}\right.
\end{equation}
with $p=1/4$. The resulting values of $\lambda$ are shown in
Fig.~\ref{fig_lambdanumerics}. The solid line is the solution of
Eq.~(\ref{eq_lambdaresult}) and the circles represent the values of
$\lambda$ for uncorrelated local scores~(\ref{eq_etaindep}). As shown
in Fig.~\ref{fig_lambdanumerics} the observed $\lambda$'s follow the
analytic solution very closely, thereby confirming
Eq.~(\ref{eq_lambdaresult}). We also included the values of $\lambda$
which result from correlated local scores generated from aligning
randomly chosen sequences according to Eq.~(\ref{eq_idmatrix}).  As
one can see, they deviate only slightly from the analytical result for
uncorrelated disorder. This deviation is strongest close to the
log-linear phase transition, which for uncorrelated disorder happens
at $\mu_{\mathrm c}=2$. The difference of $\sim2\%$ in $\mu_{\mathrm
c}$ between the correlated and the uncorrelated case rapidly becomes
much smaller for larger alphabet sizes $c$~\cite{monv99}.

\section{More general scoring systems}\label{sec_generalize}

While the approximation of the ensemble of random sequences by the ensemble
of independent local scores appears to have negligible effects, our
treatment is so far limited to the special scoring system
Eq.~(\ref{eq_etadist}).  While the computation of the generating
function $\langle\exp[-\lambda\totalhops]\rangle_0$ seems feasible
only for this special scoring system, the mapping to an asymmetric
exclusion process and the reformulation as an eigenvalue problem is
still possible for more general scoring systems~\cite{bund99b}.

We consider here scoring systems satisfying the following two
conditions: First, the differences between the possible values
$\smatrix_{\letterone,\lettertwo}$ of the scoring matrix are multiples
of some score unit $\Delta$.  Second, the gap costs $\delta$ is such
that $2\delta+s_0$ is also an integer multiple of $\Delta$, with
\begin{equation}
\smatrix_0\equiv
\max_{\letterone,\lettertwo}\{\smatrix_{\letterone,\lettertwo}\}
\end{equation}
being the maximal entry of the scoring matrix
$\smatrix_{\letterone,\lettertwo}$.  These two conditions are easily
satisfied (with $\Delta=1$) by the most frequently used protein
scoring systems~\cite{dayh78,heni92} which use integer scores and gap
costs for performance reasons.  For the match--mismatch scoring
system~(\ref{eq_idmatrix}), the first condition is satisfied with
$\Delta=1+\mu$, while the second condition applies only to a discrete
set of $\delta$'s. However, it is possible in principle to interpolate
to arbitrary gap costs~\cite{bund99}.

Mapping to an asymmetric exclusion process is possible for scoring systems
satisfying the above two conditions. It will be convenient to
express the gap cost $\delta$ in the following way,
\begin{equation}\label{eq_ndef}
2\delta=\binsize\Delta-\smatrix_0\quad\mbox{with $\binsize\in{\bf N}$}.
\end{equation}
As before, we shall ignore correlations between the local scores
$\smatrix(r,\timevar)$ and introduce uncorrelated random variables
$\disorder(r,\timevar)\in\{0,1,\ldots\}$ such that
\begin{equation}
\smatrix(r,\timevar)\equiv \smatrix_0-\disorder(r,\timevar)\Delta,
\end{equation}
i.e.,
\begin{equation}
\Pr\{\forall_{r,\timevar}\,\disorder(r,\timevar)=\disorder_{r,\timevar}\}=
\prod_{r,\timevar}\Pr\{\disorder(r,\timevar)=\disorder_{r,\timevar}\}
\end{equation}
with
\begin{equation}\label{eq_fulletadist}
\Pr\{\disorder(r,\timevar)=\disorder\}=\sum_{\letterone,\lettertwo}
p_\letterone p_\lettertwo
\delta_{\smatrix_{\letterone,\lettertwo},\smatrix_0-\disorder\Delta}.
\end{equation}
Note, that these random variables $\disorder(r,\timevar)$ only take on
a finite number of different positive integer values, since the
scoring matrix $\smatrix_{\letterone,\lettertwo}$ itself has only a
finite number of entries.

\begin{figure}[htbp]
%\narrowtext
\begin{center}
\epsfig{figure=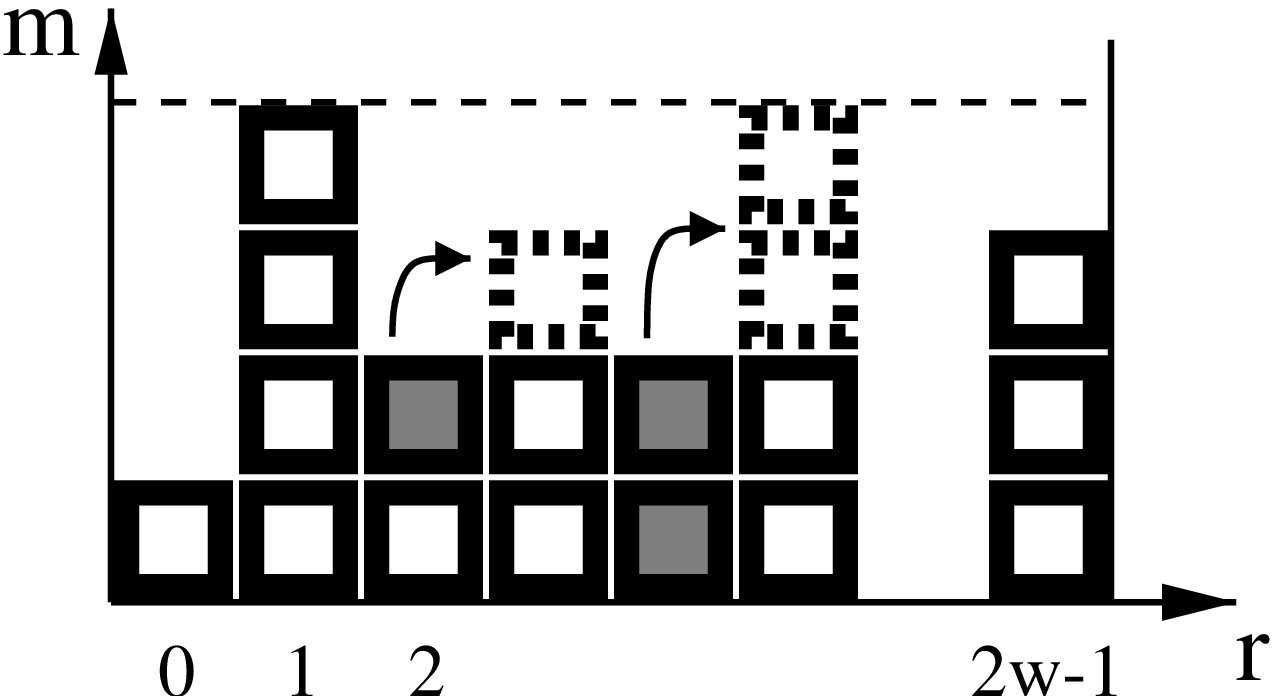,width=0.5\columnwidth}
%FIGSPACE:60mm
\caption{Interpretation of
Eqs.~(\protect\ref{eq_generalasepfirst})-(\protect\ref{eq_generalaseplast}) as
a generalized asymmetric exclusion process. A configuration of the
local score differences is represented by particles on a
one-dimensional lattice of width \protect$2\width$. Each lattice site
can accommodate up to \protect$\binsize$ particles (here
\protect$\binsize=4$.) At an odd time step for each even site
\protect$r-1$, a number of particles is chosen to attempt hopping to
the right. If there are enough particles at site \protect$r-1$ and
enough space on site \protect$r$, the chosen number hops. In the
example shown, the filled particles are the ones to hop and the dashed
boxes show their positions after the time step. No particle which
could hop is on site \protect$6$.  The particle on site \protect$0$
cannot hop since its destination site is already fully occupied. For
site \protect$2$, one particle has been chosen. On site \protect$4$, at
least two particles tried to hop. If the number chosen was larger, it
would have been cut down to two since there are only two particles on
site \protect$4$ and since there are only two spaces left at site
\protect$5$.}
\label{fig_particles}
\end{center}
\end{figure}

A derivation analogous to the one given above for the longest common
subsequence problem again maps the dynamics of the alignment algorithm
onto the dynamics of particles on a one-dimensional lattice. The state
of the system is still given by the number of particles
$\occupation(r,\timevar)$ at each lattice site, but now these
occupation numbers are defined as
\begin{equation}
\occupation(r,\timevar)\equiv\left\{
\begin{array}{ll}\frac{1}{\Delta}[\score(r+1,\timevar)-\score(r,\timevar+1)+
\delta+\smatrix_0]&
r+\timevar{\mathrm\ even}\\[5pt]
\frac{1}{\Delta}[\score(r+1,\timevar+1)-\score(r,\timevar)+\delta]&
r+\timevar{\mathrm\ odd}\end{array}\right.
\end{equation}
and can take any integer value between $0$ and $\binsize$. The
dynamics is given by the relations
\begin{eqnarray}\label{eq_generalasepfirst}
\occupation(r-1,\timevar)&=&
\occupation(r-1,\timevar-1)-\hopnumber(r,\timevar)\mbox{\ and}\\
\occupation(r,\timevar)&=&\occupation(r,\timevar-1)+\hopnumber(r,\timevar)
\end{eqnarray}
for even $r+\timevar$, where
\begin{equation}\label{eq_generalaseplast}
\hopnumber(r,\timevar)\equiv
\min\{\disorder(r,\timevar),\binsize-\occupation(r,\timevar-1),
\occupation(r-1,\timevar-1)\}
\end{equation}
and the total number of particles is fixed to be
\begin{equation}
\frac{1}{2\width}\sum_{r=0}^{2\width-1}\occupation(r,\timevar)
=\frac{\binsize}{2}.
\end{equation}
Eqs.~(\ref{eq_generalasepfirst})-(\ref{eq_generalaseplast}) can be
equally expressed as the following cellular automata: For each time
step and for each pair of neighboring sites of the one-dimensional
lattice the particles live on,
\begin{enumerate}
\item choose an integer number $\disorder\ge0$ of particles to hop
from site $r-1$ to site $r$ according to the
distribution~(\ref{eq_fulletadist})
\item if there are fewer particles than $\disorder$ on site $r-1$,
then reduce $\disorder$ to the number of particles on site $r-1$
\item if there are fewer free spaces than $\disorder$ on site $r$,
then reduce $\disorder$ to the number of free spaces on site $r$
\item move $\disorder$ particles from site $r-1$ to site $r$
\end{enumerate}
This updating rule is to be applied sublattice-parallel as for the
simpler scoring system. The process is illustrated in
Fig.~\ref{fig_particles}.

The more complicated hopping process is reflected in a different
matrix ${\sf T}_1(\lambda/\width)$ without changing anything else in
the calculations. Thus, the significance assessment constant $\lambda$
is still given by the generating function of the space and time
averaged current as
\begin{equation}\label{eq_lambdaimplicit}
\exp[\lambda s_0/2]
\langle\exp[-\lambda\Delta\totalhops]\rangle_0^{\frac{1}{N}}=1
\end{equation}
but the calculation of this generating function for an arbitrary
distribution~(\ref{eq_fulletadist}) becomes much more difficult for
the generalized asymmetric exclusion process than for the case
$\binsize=1$ of the original asymmetric exclusion process.

Already, the knowledge of the dependence of the average current on the
scoring parameters would be very helpful to biologists, since this
determines the position of the log-linear phase transition. As
discussed in the case of the simpler scoring system, the phase
transition occurs, if the first moment of the score distribution
vanishes, i.e., for
\begin{equation}
0=\left.\frac{\mathrm d}{{\mathrm d}\lambda}\right|_{\lambda=0}
\exp[\lambda s_0/2]\langle\exp[-\lambda\Delta\totalhops]\rangle_0^{\frac{1}{N}}
=s_0/2-\frac{\langle\totalhops\rangle_0}{N}\Delta=
s_0/2-\langle\hopnumber\rangle_0\Delta
\end{equation}
The average current is much easier to calculate, since in contrast to
the generating function, it is independent of temporal
correlations. Thus, it can be calculated from the knowledge of the
stationary state alone. For the original asymmetric exclusion process,
the occupation numbers of the stationary state become independent
random variables. For the generalized asymmetric exclusion process
presented here, this is not the case any more. If the number of
particles which hop in one move is at most one (as for the scoring
system~(\ref{eq_idmatrix}) with arbitrary gap costs) approximating the
stationary state as a product state still yields reasonable values of
$\langle\hopnumber\rangle_0$ and hence the phase transition point
$(\delta_{\mathrm c},\mu_{\mathrm c})$~\cite{bund99}.  Nevertheless,
exact results or at least systematic improvements taking into account
the spatial correlations of the occupation numbers would be desirable.
For the more general case allowing for an arbitrary number of
particles to hop at a given time, no analytical result is known.

\section{Concluding Remarks}

In this paper, we have shown how a question of great practical
importance to molecular biologists, like the significance assessment
of local sequence alignment results, can be answered by studying the
asymmetric exclusion process, an exactly solvable model of the KPZ
universality class. Conversely, in trying to answer this question for
biologists, we derived an important physical quantity like the
generating function $\gfgeneral$ for the corresponding physical system
in discrete time and discrete space. This complements the existing
solutions in continuous time and space~\cite{kard85} and in continuous
time and discrete space~\cite{derr98b}. Our result is the first
successful analytical approach to assessing the statistical
significance of sequence alignment with gaps.

Future work of practical importance includes solving the
generalizations of the asymmetric exclusion process described in
Sec.~\ref{sec_generalize} and studying the effect of the widely used
``affine gap cost'', where a contiguous gap of length $\ell$ is
assigned some gap cost $\delta+(\ell-1)\epsilon$ instead of simply
$\delta\ell$. A general expression which gives $\lambda$ as a function
of an arbitrary scoring system should finally give rise to a deeper
understanding of the role of the gap cost and lead to better choices
of scoring systems for alignments of biological sequences.

\section*{Acknowledgments}

The author gratefully acknowledges discussions with S. Altschul,
T. Hwa, R. Olsen, and N. Rajewsky, and the hospitality of the Center
for Studies in Physics and Biology at Rockefeller University where
this work was completed.  This work is supported in part by a
Hoch\-schul\-son\-der\-pro\-gramm~III fellowship of the DAAD and by
the NSF through Grant No. DMR-9971456.

\appendix

\section{Island High Score Distribution}\label{app_gllambda}

In this appendix we derive heuristically the Poisson distribution of
maximal island scores. We first treat the gapless case~\cite{fish99}
and then generalize the derivation to alignment with gaps.  In the
gapless case, the distribution of large islands of length $L$ measured
from their beginning to their peak point at height $\islandscore$ is
given by
\begin{equation}
p(\islandscore)=\langle\delta(\islandscore-\sum_{i=1}^L \smatrix(i))\rangle.
\end{equation}
Using the Fourier representation of the delta function and the statistical
independence of the $\smatrix(i)$ this yields
\begin{equation}
p(\islandscore)=
\frac{1}{2\pi}\int\exp(-ik\islandscore)
\langle\exp(ik\smatrix)\rangle^L{\mathrm d}k.
\end{equation}
If we assume that the peak score of the island is proportional to its
length, i.e., that an island has on average a linear slope $\alpha$, we
get
\begin{equation}
p(\islandscore)=
\frac{1}{2\pi}\int\exp(-ik\alpha L)
\langle\exp(ik\smatrix)\rangle^L{\mathrm d}k,
\end{equation}
which can be evaluated in a saddle point approximation as
\begin{equation}
p(\islandscore)\sim \exp(-\lambda \islandscore)
\end{equation}
with
\begin{equation}\label{eq_lambdasp}
\lambda=ik^*-\log[\langle\exp(ik^*\smatrix)\rangle]/\alpha.
\end{equation}
The saddle point $k^*$ is given by the saddle point equation
\begin{equation}\label{eq_saddlepoint}
\frac{\langle \smatrix\exp(ik^* \smatrix)\rangle}{
\langle\exp(ik^*\smatrix)\rangle\alpha}=1.
\end{equation}
This $k^*$ is itself a function of the so far unknown slope
$\alpha$. To find the correct value of $\alpha$, we minimize
Eq.~(\ref{eq_lambdasp}) with respect to $\alpha$ and get together with
Eq.~(\ref{eq_saddlepoint})
\begin{equation}
\langle \exp(ik^*\smatrix)\rangle=1.
\end{equation}
Inserting this into Eq.~(\ref{eq_lambdasp}) yields
condition~(\ref{eq_lambdacond}). Additionally we get from
Eq.~(\ref{eq_saddlepoint}) the typical slope $\alpha$ of an island as
\begin{equation}
\alpha=\langle\smatrix\exp(\lambda\smatrix)\rangle.
\end{equation}

For alignment with gaps, the high score of an island of length $L$
from its beginning to its peak point is not just the sum of local
scores any more.  Instead, it is given by the final score $\score(0,L)$ of
a global alignment of two sequences of length $L$ taking into account
all possible insertions of gaps. We can still use the Fourier
transformation to get
\begin{equation}
p(\islandscore)=\langle\delta(\islandscore-\score(0,L))\rangle
=\frac{1}{2\pi}\int\exp(-ik\islandscore)\langle\exp(ik\score(0,L))\rangle
{\mathrm d}k.
\end{equation}
In Sec.~\ref{sec_eigenvalue} we will see, that $\langle\exp(\lambda
\score(0,L))\rangle$ is for large $L$ the $L$'th power of the eigenvalue
of some matrix. We thus define $\eigenval(\lambda)$ by
\begin{equation}
\langle\exp[\lambda\score(0,L)]\rangle\equiv\eigenval^{L}(\lambda)
\end{equation}
and again assume a linear slope $\alpha$ of the islands which we
conveniently define by $\islandscore=\alpha L/2$ in order to take into
account that the lattice of length $L$ actually only contains $L/2$
matches or mismatches in a row. We then get
\begin{equation}
p(\islandscore)=\frac{1}{2\pi}\int\exp[(-ik\alpha/2+\log\eigenval(ik))L]
{\mathrm d}k.
\end{equation}
Applying the above saddle point approximation and maximization with
respect to the slope of the island $\alpha$ yields
Eq.~(\ref{eq_glambda}). Moreover it gives the typical slope of an
island as
\begin{equation}
\alpha=2\frac{\eigenval^\prime(\lambda)}{\eigenval(\lambda)}
=\frac{2}{L}\langle \score(0,L)\exp[\lambda \score(0,L)])\rangle.
\end{equation}

\section{Expression of the score dynamics in terms of particle occupation
numbers}\label{app_vtransform}

In this appendix we describe the mapping from the evolution
equation~(\ref{eq_lcsrecurs}) of the sequence alignment scores onto
the asymmetric exclusion process with the $\occupation(r,\timevar)$ as
the particle occupation numbers in detail.  To this end we apply
Eq.~(\ref{eq_lcsrecurs}) to the definition Eq.~(\ref{eq_occupdiff}) of
$\occupation(r,\timevar)$, where we assume by convention that
$r+\timevar$ is even as in Fig.~\ref{fig_element}(a). We get
\begin{eqnarray*}
\occupation(r\!-\!1,\timevar)\!&=&
\score(r,\timevar+1)-\score(r-1,\timevar)\\
&=&
\max\{\score(r,\timevar-1)+\disorder(r,\timevar),
\score(r-1,\timevar),\score(r+1,\timevar)\}-\score(r-1,\timevar)\}\\
&=&\score(r,\timevar\!-\!1\!)\!-\!\score(r\!-\!1,\timevar)\!+\!1
\!+\!\max\{\disorder(r,\timevar)\!-\!1,
\score(r\!-\!1,\timevar)\!-\!\score(r,\timevar\!-\!1\!)\!-\!1,
\score(r\!+\!1,\timevar)\!-\!\score(r,\timevar\!-\!1\!)\!\!-\!1\}\\
&=&\occupation(r-1,\timevar-1)+
\max\{\disorder(r,\timevar)\!-\!1,
-\occupation(r\!-\!1,\timevar\!-\!1),
\occupation(r,\timevar\!-\!1)\!-\!1\}\\
&=&\occupation(r-1,\timevar-1)-
\min\{1\!-\!\disorder(r,\timevar),\occupation(r\!-\!1,\timevar\!-\!1),
1-\occupation(r,\timevar\!-\!1)\}
\end{eqnarray*}
and analogously
\begin{eqnarray*}
\occupation(r,\timevar)\!&=&
\score(r+1,\timevar)-\score(r,\timevar+1)+1\\
&=&\score(r+1,\timevar)-
\max\{\score(r,\timevar-1)+\disorder(r,\timevar),
\score(r-1,\timevar),\score(r+1,\timevar)\}+1\\
&=&\score(r\!+\!1,\timevar)\!-\!\score(r,\timevar\!-\!1)
\!-\!\max\{\disorder(r,\timevar)\!-\!1,
\score(r\!-\!1,\timevar)\!-\!\score(r,\timevar\!-\!1\!)\!-\!1,
\score(r\!+\!1,\timevar)\!-\!\score(r,\timevar\!-\!1\!)\!\!-\!1\}\\
&=&\occupation(r,\timevar-1)-
\max\{\disorder(r,\timevar)\!-\!1,
-\occupation(r\!-\!1,\timevar\!-\!1),
\occupation(r,\timevar\!-\!1)\!-\!1\}\\
&=&\occupation(r,\timevar-1)+
\min\{1\!-\!\disorder(r,\timevar),\occupation(r\!-\!1,\timevar\!-\!1),
1-\occupation(r,\timevar\!-\!1)\}.
\end{eqnarray*}
This can be summarized in the form
\begin{eqnarray}\label{eq_asepdynfirst}
\occupation(r-1,\timevar)&=&
\occupation(r-1,\timevar-1)-\hopnumber(r,\timevar)\quad\mbox{and}
\\\label{eq_asepdynr}
\occupation(r,\timevar)&=&\occupation(r,\timevar-1)+\hopnumber(r,\timevar),
\end{eqnarray}
where
\begin{equation}\label{eq_asepdynlast}
\hopnumber(r,\timevar)\equiv
\min\{1-\disorder(r,\timevar),1-\occupation(r,\timevar-1),
\occupation(r-1,\timevar-1)\}.
\end{equation}

As we can see, there is no reference to the actual alignment scores
$\score(r,\timevar)$ in these equations.  As a first consequence of
these equations we note that they imply that the variables
$\occupation(r,\timevar)$ can only take on the values zero and one.
This is obvious by induction, if it is fulfilled at $\timevar=0$ as it
is the case for our choice of initial conditions\footnote{Even if the
initial values of the $\occupation(r,\timevar=0)$ are not zero or one
they will under the dynamics
Eqs.~(\ref{eq_asepdynfirst})-(\ref{eq_asepdynlast}) eventually try to
take on values less than zero or larger than one.  The minimum in
Eq.~(\ref{eq_asepdynlast}) then resets them to zero or one. Thus,
after some startup phase, the $\occupation(r,t)$ will be integer even
if their initial values are chosen to be non-integer.}.  Thus, it is
reasonable to interpret the $\occupation(r,\timevar)$ as particle
occupation numbers.

Moreover, we note that a pair of neighboring occupation numbers
$(\occupation(r-1,\timevar),\occupation(r,\timevar))$ at time $\timevar$
depends only on the corresponding pair
$(\occupation(r-1,\timevar-1),\occupation(r,\timevar-1))$ at time
$\timevar-1$ and the random variable $\disorder(r,\timevar)$. Thus, the
elements as the one shown in Fig.~\ref{fig_element} perform these
transformations of a pair of neighboring occupation numbers into a
new pair of neighboring occupation numbers completely independently
from each other.

Looking at Eqs.~(\ref{eq_asepdynfirst})-(\ref{eq_asepdynlast}) more
closely, we see that $\hopnumber(r,\timevar)=0$ whenever
$(\occupation(r-1,\timevar-1),\occupation(r,\timevar-1))\in\{|00\rangle,
|01\rangle,|11\rangle$\}. Thus,
$(\occupation(r-1,\timevar),\occupation(r,\timevar))=
(\occupation(r-1,\timevar-1),\occupation(r,\timevar-1))$ in these
cases.  Only if site $r-1$ is occupied and site $r$ is empty, the
number $\hopnumber(r,\timevar)$ of transfered particles can be one
with probability $\Pr\{\disorder(r,\timevar)=0\}=1-p$. This leads to
the interpretation of the dynamics given by
Eqs.~(\ref{eq_asepdynfirst})-(\ref{eq_asepdynlast}) as an asymmetric
exclusion process described by the transfer matrix ${\sf T}_1(0)$ defined
in Eq.~(\ref{eq_transfermatrix}) of the main text.

So far we transformed the dynamics of the sequence alignment algorithm
as given by Eq.~(\ref{eq_lcsrecurs}) into an asymmetric exclusion
process. We still have to express $\gfgeneral_0(\lambda;N)$ in terms of this
asymmetric exclusion process. To achieve this, we first
define for any ``time'' $\timevar$ the average score (or
space-averaged surface height)
\begin{equation}
\overline{\score}(\timevar)\equiv\left\{
\begin{array}{ll}
\frac{1}{2\width}\sum_{k=0}^{\width-1}
[\score(2k,\timevar-1)+\score(2k+1,\timevar)]&
\timevar{\mathrm\ even}\\[5pt]
\frac{1}{2\width}\sum_{k=0}^{\width-1}
[\score(2k,\timevar)+\score(2k+1,\timevar-1)]&
\timevar{\mathrm\ odd}
\end{array}\right.
\end{equation}
Because of the translational invariance of the system in the spatial ($r$)
direction we get
\begin{equation}\label{eq_sbysbar}
\gfgeneral_0(\lambda;N)=\langle\exp[\lambda\score(0,N)]\rangle_0=
\langle\exp[\lambda\overline{\score}(N)]\rangle_0.
\end{equation}
Thus, we can restrict ourselves to calculating the large $N$ behavior
of the latter quantity.

The change in the average score $\overline{\score}(\timevar)$ can be
expressed in terms of the occupation numbers $\occupation(r,\timevar)$
via Eqs.~(\ref{eq_lcsrecurs}) and~(\ref{eq_occupdiff}). It is given by
\begin{equation}\label{eq_scoreadvancebyscorediff}
\overline{\score}(\timevar+1)-\overline{\score}(\timevar)
=\left\{\begin{array}{ll}
\frac{1}{2\width}\sum_{k=0}^{\width-1}
[\score(2k,\timevar+1)-\score(2k,\timevar-1)]&\timevar{\mathrm\ even}\\[5pt]
\frac{1}{2\width}\sum_{k=0}^{\width-1}
[\score(2k+1,\timevar+1)-\score(2k+1,\timevar-1)]&\timevar{\mathrm\ odd}
\end{array}\right..
\end{equation}
The local score differences in this equation can for even $r+\timevar$
be expressed as
\begin{eqnarray*}
\score(r,\timevar+1)-\score(r,\timevar-1)
&=&\max\{\score(r,\timevar-1)+\disorder(r,\timevar),
\score(r+1,\timevar),\score(r-1,\timevar)\}-\score(r,\timevar-1)\\
&=&1+\max\{\disorder(r,\timevar)-1,
\occupation(r,\timevar-1)-1-\occupation(r-1,\timevar-1)\}\\
&=&1-\min\{1-\disorder(r,\timevar),
1-\occupation(r,\timevar-1),\occupation(r-1,\timevar-1)\}\\
&=&1-\hopnumber(r,\timevar).
\end{eqnarray*}
Inserting this into Eq.~(\ref{eq_scoreadvancebyscorediff}) yields
\begin{equation}\label{eq_shataslocal}
\overline{\score}(\timevar+1)-\overline{\score}(\timevar)
=\frac{1}{2}-\frac{1}{2\width}\left\{\begin{array}{ll}
\sum_{k=0}^{\width-1}\hopnumber(2k,\timevar)&\timevar{\mathrm\ even}\\[5pt]
\sum_{k=0}^{\width-1}\hopnumber(2k+1,\timevar)&\timevar{\mathrm\ odd}
\end{array}\right..
\end{equation}

Combining Eqs.~(\ref{eq_sbysbar}) and~(\ref{eq_shataslocal}) finally
yields
\begin{eqnarray}
\gfgeneral_0(\lambda;N)&=&
\langle\exp[\lambda \overline{\score}(N)]\rangle_0=
\langle
\exp[\lambda\sum_{\timevar=0}^{N-1}
(\overline{\score}(\timevar+1)-\overline{\score}(\timevar))]\rangle_0
\\\nonumber
&=&\exp[\lambda N]
\langle\exp[-\frac{\lambda}{2\width}
\sum_{l=1}^{N/2}\sum_{k=0}^{\width-1}
(\hopnumber(2k+1,2l-1)+\hopnumber(2k,2l))]\rangle_0\\\nonumber
&=&\exp[\lambda N]\langle\exp[-\lambda\totalhops]\rangle_0,
\end{eqnarray}
where
\begin{equation}
\totalhops\equiv\frac{1}{2\width}\sum_{l=1}^{N/2}\sum_{k=0}^{\width-1}
(\hopnumber(2k+1,2l-1)+\hopnumber(2k,2l))
\end{equation}
is the total number of particles hopped divided by the number of
sites. This is Eq.~(\ref{eq_genfuncsbyeachother}) of the main text.

\section{Dynamic Path Integral Representation}\label{app_pathintegral}

In this appendix we want to show that the generating function
$\genfunc(\lambda;\width,N)$ can be expressed as a product of some
$4^{\width}$ dimensional matrices as stated in
Eq.~(\ref{eq_genasmatrixproduct}) in the main text.  This rewriting is
crucial in transforming the calculation of the generating function
into an eigenvalue problem. We start from the definition
\begin{equation}\label{eq_elsinpieces}
\genfunc(\lambda;\width,N)=\langle\exp[-\lambda\totalhops]\rangle_0=
\langle\prod_{l=1}^{N/2}
\prod_{k=0}^{\width-1}e^{-\frac{\lambda}{2\width}\hopnumber(2k+1,2l-1)}
e^{-\frac{\lambda}{2\width}\hopnumber(2k,2l)}\rangle_0.
\end{equation}
Since, the number of particles in each bin must be either $0$ or $1$
at any time, we do not change the expectation value, if we introduce ones
of the form
\begin{equation}
1=\sum_{\{\occupation_{r,\timevar}\}\in\{0,1\}^{2\width}}
\prod_{r=0}^{2\width-1}
\delta_{\occupation(r,\timevar),\occupation_{r,\timevar}}
\end{equation}
at each fixed time $\timevar$. This corresponds to a path integral formulation
of the quantity $\genfunc(\lambda;\width,N)$ and yields
\begin{eqnarray}\label{eq_genassum}
\lefteqn{\langle\exp[-\lambda\totalhops]\rangle_0=}\hspace{10mm}&&\\\nonumber
&&\sum_{\{\occupation_{r,0}\}}\ldots\sum_{\{\occupation_{r,N}\}}
\langle
\prod_{r=0}^{2\width-1}\delta_{\occupation(r,0),\occupation_{r,0}}
\prod_{l=1}^{N/2}
\left(\prod_{r=0}^{2\width-1}
\delta_{\occupation(r,2l-1),\occupation_{r,2l-1}}\right)
\left(\prod_{k=0}^{\width-1}
e^{-\frac{\lambda}{2\width}\hopnumber(2k+1,2l-1)}\right)
\times\\\nonumber&&\hspace{5.8cm}\times
\left(\prod_{r=0}^{2\width-1}
\delta_{\occupation(r,2l),\occupation_{r,2l}}\right)
\left(\prod_{k=0}^{\width-1}
e^{-\frac{\lambda}{2\width}\hopnumber(2k,2l)}\right)\rangle_0
\end{eqnarray}
Once a configuration of the particles at each time step is fixed, the
expectation value can be factorized into the parts which contain only
a single random variable $\disorder(r,\timevar)$
\begin{eqnarray*}
\lefteqn{\langle
\prod_{r=0}^{2\width-1}\delta_{\occupation(r,0),\occupation_{r,0}}
\prod_{l=1}^{N/2}
\left(\prod_{r=0}^{2\width-1}
\delta_{\occupation(r,2l-1),\occupation_{r,2z-1}}\right)
\left(\prod_{k=0}^{\width-1}
e^{-\frac{\lambda}{2\width}\hopnumber(2k+1,2l-1)}\right)
\left(\prod_{r=0}^{2\width-1}
\delta_{\occupation(r,2l),\occupation_{r,2l}}\right)
\left(\prod_{l=0}^{\width-1}
e^{-\frac{\lambda}{2\width}\hopnumber(2k,2l)}\right)
\rangle_0=}\hspace{5mm}&&\\
&&\prod_{r=0}^{2\width-1}\delta_{\occupation(r,0),\occupation_{r,0}}\times\\
&&\!\times\!\prod_{l=1}^{N/2}
\prod_{k=0}^{\width-1}
\langle\delta_{\occupation(2k,2l-2),\occupation_{2k,2l-2}}
\delta_{\occupation(2k+1,2l-2),\occupation_{2k+1,2l-2}}
e^{-\frac{\lambda}{2\width}\hopnumber(2k+1,2l-1)}
\delta_{\occupation(2k,2l-1),\occupation+{2k,2l-1}}
\delta_{\occupation(2k,2l-1),\occupation_{2k+1,2l-1}}\rangle_0\!\times\!\\
&&\qquad\times\prod_{k=0}^{\width-1}
\langle\delta_{\occupation(2k-1,2l-1),\occupation_{2k-1,2l-1}}
\delta_{\occupation(2k,2l-1),\occupation_{2k,2l-1}}
e^{-\frac{\lambda}{2\width}\hopnumber(2k,2l)}
\delta_{\occupation(2k-1,2l),\occupation_{2k-1,2l}}
\delta_{\occupation(2k,2l),\occupation_{2k+1,2z}}\rangle_0\times1.
\end{eqnarray*}
Inserting this into Eq.~(\ref{eq_genassum}) we can interpret the
summation over the possible configurations of the particles at each
time step as the summation of inner indices in a matrix
multiplication. In this language the first term
$\prod_{r=0}^{2\width-1}\delta_{\occupation(r,0),\occupation_{r,0}}$
is a vector on the $4^{\width}$ dimensional vector space indexed by
all possible particle configurations. This vector has exactly one non
vanishing entry at the configuration which is chosen as the initial
configuration at $\timevar=0$.  This non vanishing entry is one and we
call this vector $|\psi_0\rangle$. The factor of one which we added for
the sake of clarity also plays the role of a vector the entries of
which are all one.  We call this vector $\langle\psi_1|$. All the other
factors represent matrices.  There is one matrix for every time step
and each of these matrices is a tensor product of $\width$ identical
matrices describing an elementary hopping process.  Their matrix
elements are
\begin{equation}
\left({\sf T}_1\left(\frac{\lambda}{\width}\right)
\right)_{(\occupation_1,\occupation_2),
(\occupation_1^\prime,\occupation_2^\prime)}\equiv
\langle
\delta_{\occupation(r-1,\timevar-1),\occupation_1^\prime}
\delta_{\occupation(r,\timevar-1),\occupation_2^\prime}
\exp[-\frac{\lambda}{2\width}\hopnumber(r,\timevar)]
\delta_{\occupation(r-1,\timevar),\occupation_1}
\delta_{\occupation(r,\timevar),\occupation_2}\rangle_0.
\end{equation}
The disorder average here is over one single random variable
$\disorder(r,\timevar)$. Performing this disorder average yields the
matrix ${\sf T}_1(\lambda/\width)$ as defined in
Eq.~(\ref{eq_deftoflambda}).  The matrices for even time steps and the
matrices for odd time steps are shifted against each other by one
lattice unit which finally leads to the expression of
Eq.~(\ref{eq_genasmatrixproduct}).

\end{document}